\documentclass[pra,twocolumn,superscriptaddress,showpacs,amsmath,amssymb]{revtex4-2}

\usepackage{graphicx}
\usepackage{dcolumn}
\usepackage{bm}
\usepackage{epsfig}
\usepackage{amsmath}
\usepackage{amssymb}
\usepackage{color}
\usepackage[dvipsnames]{xcolor}
\usepackage{bbm}
\usepackage{bbold}
\usepackage{CJK}
\usepackage[colorlinks=true,citecolor=blue,linkcolor=blue,urlcolor=blue]{hyperref}
\begin{document}
\title{Simulating monitoring-induced topological phase transitions with small systems}
\author{Rui Xie}
\affiliation{Zhejiang Key Laboratory of Quantum State Control and Optical Field Manipulation, Department of Physics, Zhejiang Sci-Tech University, 310018 Hangzhou, China}
\author{Clemens Gneiting}
\affiliation{Theoretical Quantum Physics Laboratory, Cluster for Pioneering Research, RIKEN, Wakoshi, Saitama 351-0198, Japan}
\affiliation{Center for Quantum Computing, RIKEN, Wakoshi, Saitama 351-0198, Japan}
\author{Zheng-Yang Zhou}
\altaffiliation[zheng-yang.zhou@zstu.edu.cn]{}
\affiliation{Zhejiang Key Laboratory of Quantum State Control and Optical Field Manipulation, Department of Physics, Zhejiang Sci-Tech University, 310018 Hangzhou, China}
\affiliation{Theoretical Quantum Physics Laboratory, Cluster for Pioneering Research, RIKEN, Wakoshi, Saitama 351-0198, Japan}
\author{Ai-Xi Chen}
\altaffiliation[aixichen@zstu.edu.cn]{}
\affiliation{Zhejiang Key Laboratory of Quantum State Control and Optical Field Manipulation, Department of Physics, Zhejiang Sci-Tech University, 310018 Hangzhou, China}
\date{\today}

\begin{abstract}
 
 The topological properties of open quantum lattice systems have attracted much attention, due to their fundamental significance and potential applications. However, experimental demonstrations with large-scale lattice models remain challenging. On top of that, formulations of topology in terms of quantum trajectories require monitoring along with the detection of quantum jumps. This is particularly the case for the dark state-induced topology that relies on averaging quantum trajectories at their jump times. Here, we propose two significant simplifications to ease the experimental burden to demonstrate dark-state induced topological phase transitions: First, we emulate the topology in the phase space of small systems, where the effective size of the system is reflected by the accessible parameter range. Second, we develop a method how to, by augmenting the system with an auxiliary system, access the jump-time averaged state through standard wall-time averaging, which effectively substitutes the monitoring along with the counting of quantum jumps. While these simplifications are applicable to general lattice systems, we demonstrate them with a one-dimensional Su-Schrieeffer-Heeger model. In this case, the lattice system is emulated by a four-level system, while the jump-time averaged state up to the second jump is accessed through a three-level auxiliary system.
\end{abstract}

\maketitle


%
%


\section{Introduction}
Topological properties can have interesting and far-reaching manifestations in various quantum systems~\cite{RevModPhys.82.3045, RevModPhys.83.1057, Ryu_2010,asboth2016short}. This is well established for topological insulators, where the presence of edge states can be traced back to the underlying bulk topology. Beyond topological insulators, which are closed systems, topology can also play an important role in open quantum systems. For instance, if the presence of an environment is described by a quantum master equation, topological properties can be identified in the steady-state manifold~\cite{Bardyn_2013, PhysRevX.8.011035, PhysRevLett.124.040401, PhysRevA.98.013628, PhysRevA.100.062131, PhysRevLett.123.170401, PhysRevResearch.2.033428}. Alternatively, systems that are monitored, while the occurrence of quantum jumps is ruled out through postselection, can be characterized by non-Hermitian Hamiltonians, enabling the exploration of associated topological phenomena~\cite{PhysRevLett.102.065703, PhysRevLett.115.200402, PhysRevLett.118.040401, PhysRevA.96.053858, PhysRevLett.118.200401, PhysRevLett.121.086803, PhysRevLett.121.026808, PhysRevX.8.031079, PhysRevA.99.063834, PhysRevX.9.041015, PhysRevLett.124.056802, ashida2020non}. Finally, an open system that is unraveled but quantum jumps are taken into account---thus avoiding postselection---can give access to otherwise hidden topological phenomena if the quantum trajectories are ensemble-averaged at fixed jump counts, that is, at the {\it jump times}~\cite{tra1,tra2,top2}. In view of this variety of perspectives, it is important to identify and experimentally verify the full spectrum of observable consequences that can emerge from the topological properties of open quantum systems.

As topology refers to the global properties of a system, we usually need a comparatively large system in order to control and suppress finite-size effects. As a result, the requirements on experimental samples are typically demanding, and demonstrations of topological properties are limited to large and complex platforms such as photonic networks~\cite{Lu2014, RevModPhys.91.015006, Ghatak_2019, 10.1063/1.5086433, 10.1063/1.5142397, Rudner2020, doi:10.1126/science.abf6568}, acoustic lattices~\cite{PhysRevLett.114.114301, Xue2022, Huber2016, Ni2019,Ma2019}, cold atoms~\cite{Wintersperger2020, PhysRevLett.124.250402, doi:10.1126/science.abi8794}, or superconducting circuit arrays~\cite{PhysRevB.99.020304, Gilbert2021, PhysRevX.11.011015}. Here, we circumvent the necessity of large systems by exploiting that, since topological properties are formulated in the phase space of quantum systems, they can be emulated by smaller systems, as long as these provide access to the same phase space. This equally applies to open quantum systems, where the presence of an environment in general provides access to an even larger phase space. Specifically, we elaborate this for a topologoical phase transition that occurs under the jump-time averaging~\cite{tra1,tra2,top2} of quantum trajectories~\cite{PhysRevLett.52.1657, DIOSI1986451, Belavkin1990, tra1, PhysRevA.46.4363, PhysRevLett.68.580, NGisin_1992, RevModPhys.70.101}, where we show that the topological phase-space signature originally occurring in a one-dimensional Su-Schrieffer-Heeger (SSH) model \cite{top1} can be emulated by just two quantum bits.


Observing quantum trajectories requires continuous measurements~\cite{PhysRevLett.56.2797, PhysRevLett.57.1696, PhysRevLett.57.1699, PhysRevLett.83.1287, PhysRevLett.96.076605, doi:10.1126/science.1126788, Gleyzes2007, Kubanek2009, doi:10.1126/science.1189075, PhysRevLett.106.110502, Sayrin2011, Pla2013, Minev2019, PhysRevLett.122.247403}, which in themselves can be experimentally challenging. This raises the question if there exists a way to get access to jump-time averaged states without resorting to monitoring. We develop such a method, which is based on augmenting the system by an auxiliary system. The latter, by design, autonomously keeps track of the quantum jumps that occur in the system. Measuring the auxiliary system at fixed times then allows to reconstruct the jump-time averaged system state without monitoring.

By combining these two simplifications, we propose a method to demonstrate jump-time topological effects with small, unmonitored systems. We show that the required lattice model can be emulated by a four-level system with tunable parameters. Moreover, we show that a three-level auxiliary system is sufficient to autonomously categorize different quantum trajectories by their jump count up to the second quantum jump. We prove that the respective density matrices in the jump-time domain can be reconstructed by measuring the auxiliary mode at fixed times. Following this, numerical simulations demonstrate the validity of our proposal. At the same time, viable parameters for the system and the measurements, which can also affect the required experimental resources, are identified. Finally, we discuss a possible experimental realization along with potential simplifications of our model.

This paper is structured as follows: In Sec.~\ref{Stheory}, we introduce the theoretical background and provide the dissipative SSH model that underlies our proposal. In Sec.~\ref{secnumsim}, we implement the numerical simulations, engaging in a discussion that encompasses both ideal and realistic cases. Visual representations corresponding to these cases are plotted for clarity. Section~\ref{secexppro} delves into the description of a possible experimental implementation. The conclusions are presented in Sec.~\ref{conclusion}.

\section{THEORETICAL MODEL}\label{Stheory}

\subsection{QUANTUM TRAJECTORIES} 

The time evolution of open quantum systems is often governed by a Lindblad master equation~\cite{master1, master2} (setting $\hbar=1$),
\begin{equation} \label{Eq:Lindblad_master_equation}
\begin{split}
&{\partial }_{t}{\rho  }_{t}=-i[H,{\rho  }_{t}]+\sum_{j}\gamma_j \left(L_{j}{\rho  }_{t}L_{j}^\dagger-1/2\left\{L_{j}^{\dagger}L_{j}, {\rho  }_{t} \right\}\right),\\
\end{split}
\end{equation}
where $H$ is the Hamiltonian of the closed system. For instance, Eq.~(\ref{Eq:Lindblad_master_equation}) can be derived from a microscopic model of the environment if the Born-Markov approximation is applicable. The Lindblad operators ${L}_{j}$ then describe the dissipation dynamics induced by the environment, $\left(\left \{\widehat{A},\widehat{B}\right\} = \widehat{A}\widehat{B} + \widehat{B}\widehat{A}\right)$ is the anti-commutator, and $\gamma_j$ are the environmental dissipation rates for different Lindblad operators. Alternatively, the Lindblad operators can emerge from a continuous monitoring process, which provides a physical basis for the quantum trajectories that we discuss in the following.

The Markovian master equation~(\ref{Eq:Lindblad_master_equation}) can be formally divided into non-Hermitian evolution terms and the quantum jump terms,
\begin{eqnarray}
{\partial }_{t}{\rho  }_{t}&=&-iH_{\rm eff}\rho_t+i\rho_tH_{\rm eff}^{\dag}+\sum_{j}\gamma_jL_j\rho_tL_j^{\dag},
\end{eqnarray}
with the non-Hermitian Hamiltonian $H_{\rm eff}=H-i\sum_j\gamma_j/2L^{\dag}_jL_j$ and the quantum jump terms $L_j\rho_tL_j^{\dag}$. According to the order of the jump terms, the density matrix in the time domain $\rho _{t}$ can be formally decomposed into trajectories interspersed by stochastic quantum jumps~\cite{tra1} (see also Appendix \ref{a1}),
\begin{eqnarray} \label{Eq:formal_Lindblad_solution}
&&\rho _{t}=\sum_{n=0}^{\infty} \int_{0}^{t} dt_{n}\int_{0}^{t_{n}} dt_{n-1}\dots \int_{0}^{t_{2}} dt_{1}\sum _{j_{1}\dots j_{n}\in I}\rho^{t} _{j_{n}\dots j_{1}}(\left\{t_{i} \right\}),\nonumber\\
\end{eqnarray} 
with
\begin{eqnarray}
\rho^{t} _{j_{n}\dots j_{1}}(\left\{t_{i} \right\})&\equiv&\mathcal{U}_{t-t_{n}}\mathcal{J}_{j_{n}}\mathcal{U}_{t_{n}-t_{n-1}}\mathcal{J}_{j_{n-1}}\dots \mathcal{J}_{j_{1}}\mathcal{U}_{t_{1}}\rho _{0},\nonumber\\
\mathcal{U}_{t}A&\equiv& e^{-iH_{\rm eff}t}Ae^{iH_{\rm eff}t},\nonumber\\
\mathcal{J}_{j}A&\equiv& L_jAL^{\dag}_{j},
\end{eqnarray}
where $\rho^{t} _{j_{n}\dots j_{1}}(\left\{t_{i} \right\})$ describes a (not normalized) quantum trajectory with $n$ quantum jumps at time points $t_1,t_2,\dots,t_n$. The non-Hermitian evolution and the quantum jumps are described by the superoperators $\mathcal{U}_{t}$ and $\mathcal{J}_{J_n}$, respectively. The serial numbers $J_n$, out of the set $I=\{1,2...N\}$, indicate which jump kind  is realized in the $n$th jump. 

Quantum trajectories obtain physical relevance if the jump operators are induced by continuous measurements, where the jump types $j_k$ and jump times $t_k$ correspond to the measurement records that condition the quantum trajectories. In this interpretation, the master equation (\ref{Eq:Lindblad_master_equation}) describes the dynamics of a density matrix which is obtained from ensemble averaging over all quantum trajectories at fixed times while discarding the measurement records.

It has been shown that, based on quantum trajectories, one can equally isolate density matrices corresponding to fixed jump times~\cite{tra2},
\begin{equation} \label{Eq:jumptime-averaged_state}
\begin{split}
&\rho _{n}= \int_{0}^{\infty} dt_{n}\int_{0}^{t_{n}} dt_{n-1}\dots 
 \int_{0}^{t_{2}} dt_{1}\sum _{j_{1}\dots j_{n}\in I}\rho^{n} _{j_{n}\dots j_{1}}(\left\{t_{i} \right\}),`\\
\end{split}
\end{equation} 
where $\rho _{n}$ represents the normalized density matrix associated with $n$ times of jumps, with the (unnormalized) quantum trajectories,
\begin{equation}
\rho^{n} _{j_{n}\dots j_{1}}(\left\{t_{i} \right\})=\mathcal{J}_{j_{n}}\mathcal{U}_{t_{n}-t_{n-1}}\mathcal{J}_{j_{n-1}}\dots \mathcal{J}_{j_{1}}\mathcal{U}_{t_{1}}\rho _{0},
\end{equation}
In the monitoring approach to open quantum systems, access to the state (\ref{Eq:jumptime-averaged_state}) is obtained through counting the quantum jump occurrences.

Such jump-time averaged density matrices (\ref{Eq:jumptime-averaged_state}) ``evolve'', in analogy to the $t$ dependence of the density matrices in the time domain, with the number of jumps,
\begin{eqnarray}\label{jump-timeevolution}
\rho_{n+1}=\int_0^{\infty}ds\sum_{j}\gamma_{j}\mathcal{J}_{j}\mathcal{U}_{s}\rho_n.
\end{eqnarray}
Note that the density matrices in both the time domain and the jump-time domain share the same initial state $\rho_0$.

Dark states play an outstanding role for the jump-time evolution (\ref{jump-timeevolution}). Dark states $|\rm{D}\rangle$ are eigenstates of the Hamiltonian that are annihilated by all jump operators,
\begin{align}
[H, |{\rm D}\rangle \langle {\rm D}|] = 0 \hspace{3mm} {\rm and} \hspace{3mm} L_j |{\rm D} \rangle = 0 \hspace{2mm} \forall j .
\end{align}
In the presence of dark states, the jump-time evolution Eq.~(\ref{jump-timeevolution}) may not be trace-preserving, as dark states are mapped as
\begin{eqnarray}
\rho_{n+1}&=&\int_0^{\infty}ds\sum_{j}\gamma_{j}\mathcal{J}_{j}\mathcal{U}_{s}|\rm{D}\rangle\langle \rm{D}|\nonumber\\
          &=&0 .
\end{eqnarray}
This sensitivity of the jump-time evolution (\ref{jump-timeevolution}) to the presence of dark states gives rise to the dark state-induced topology that is elaborated in the following subsection.

As the jump-time evolution from $n$ to $n+1$ jumps in Eq.~(\ref{jump-timeevolution}) does not change with the jump time, we can define a jump-time propagator $K$ independent of $n$,
\begin{eqnarray}\label{defjumptimeevolutionoperator}
\langle p|\rho_{n+1}|p'\rangle=\sum_{q,q'}K(p,q,p',q')\langle q|\rho_{n}|q'\rangle,
\end{eqnarray}
where $|p\rangle$ corresponds to a set of basis states. This propagator provides the basis for accessing the topological behavior, which will be discussed in the following part. 

\subsection{TOPOLOGY AND THE HAMILTONIAN OF THE SYSTEM}

Several topological properties of dissipative quantum systems can be revealed by the jump-time dynamics (\ref{jump-timeevolution}), based on its sensitivity on the presence of dark states~\cite{top1,top2}. Specifically, a topological order parameter can be formulated in terms of the jump-time propagator, and different topological equivalence classes of the dissipative system can then be defined based on the properties of the jump-time propagator. The corresponding topological phase transition can be observed by monitoring the jump-time averaged state of the system. Thus we can use the jump-time propagator as an indicator for the lattice topological properties of such systems. 

\begin{figure}[t]
\centering
\includegraphics[width=0.48\textwidth]{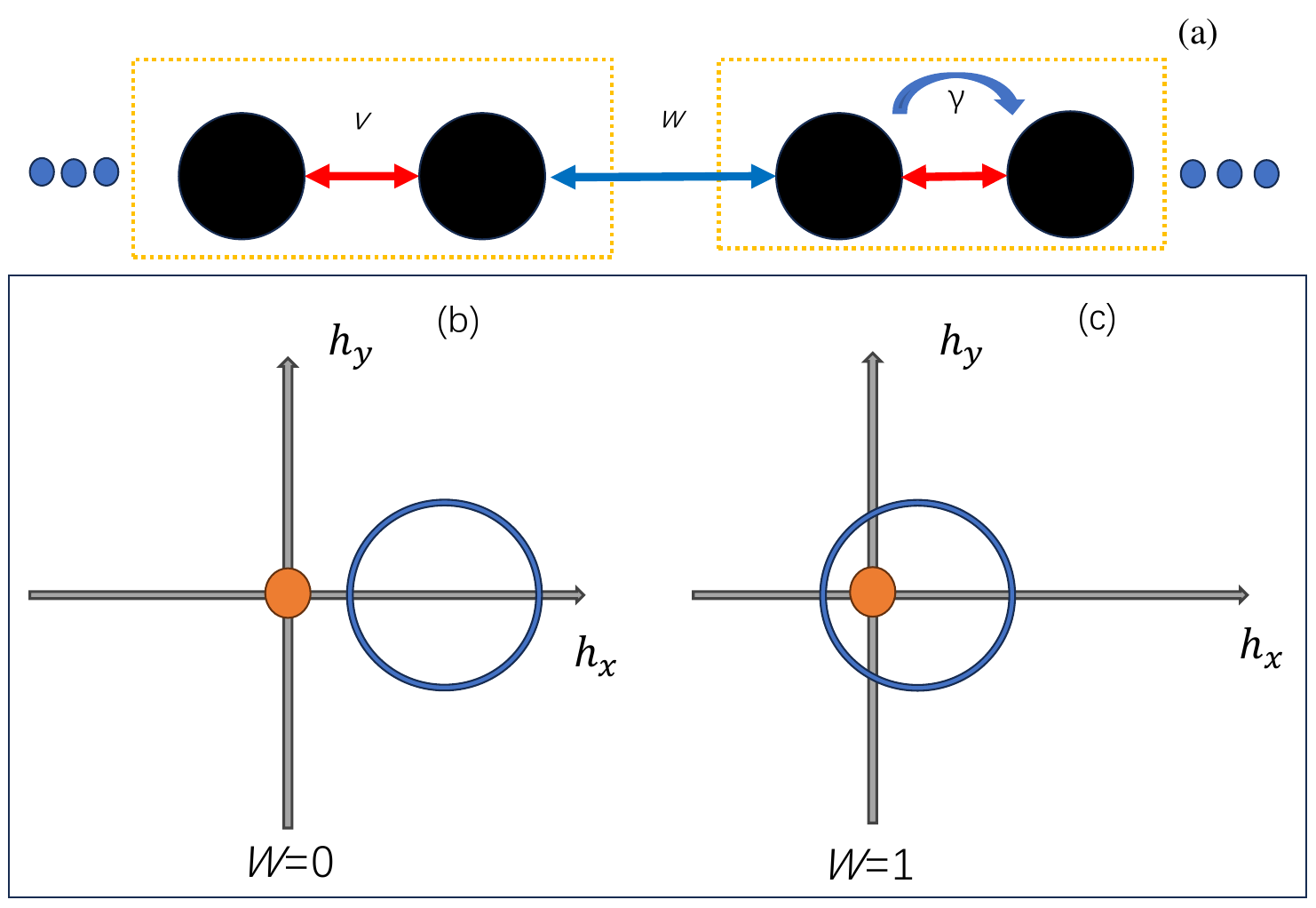} \caption{Dissipative Su-Schrieffer-Heeger model and associated topology. (a) A one-dimensional lattice with staggered hoppings, intracell hopping $v$ (red double arrows) and intercell hopping $w$ (blue curved arrow), is extended by dissipative, unidirectional intracell jumps. (b, c) The associated Bloch Hamiltonian describes a circle in the phase space, which, depending on the relation of $v$ and $w$, either (b) encircles the dark state-induced singular point at the origin or (a) does not. The corresponding winding number, (b) $W=0$ and (c) $W=1$, respectively, coincides with the jump-time phase, cf.~(\ref{tk}), and thus can be retrieved from the jump-time averaged states.}
\label{ssh}
\end{figure}

We consider a Su-Schrieffer-Heeger model~\cite{top1} with additional dissipation, as illustrated by Fig.~\ref{ssh}. The system is a lattice with two sites per unit cell. The inter-unit-cell coupling $w$ and the intra-unit-cell coupling $v$ in general take different values. In addition, we assume a collective (that is, acting on all unit cells simultaneously) decay from one sublattice to the other, which can, for instance, be caused by a common field coupled to all unit cells. Such a system can be partially diagonalized in momentum space, where the overall state space consists of the direct product of different momentum states $|p\rangle$ and an intrinsic two-level system~\cite{top2}: 
\begin{eqnarray}\label{topologicalhamiltonian}
H&=&\oint dp\left | p  \right \rangle \left \langle p\right| \otimes H(p),\nonumber\\
H(p)&=&h_x(p)\sigma_x+h_y(p)\sigma_y,\nonumber\\
h_x(p)&=&v+w\cos (p),~h_y(p)=-w\sin (p) .
\end{eqnarray}
The dissipation channel is expressed by the collective collapse Lindblad operator
\begin{eqnarray}\label{topologicaljumpoperator}
L_{\rm cc}&=&I_{p}\otimes \sigma_-,
\end{eqnarray}
where $\sigma_k$ $(k=x,y,z)$ are the Pauli matrices, $\sigma_-=\sigma_x-i\sigma_y$, and $I_{p}$ denotes the identity operator in momentum space. 
To simplify notation, we set $v=1$ and express other parameters as ratios of $v$. We emphasize that the restriction to a vanishing $z$ component in the Hamiltonian (\ref{topologicalhamiltonian}), which reflects the chiral symmetry of the SSH Hamiltonian, is not essential to observe the topological phase transition of the jump-time unraveled system. Indeed, it has been shown that non-vanishing $z$ components can be admitted given they are momentum-independent or given the Hamiltonian satisfies time-reversal symmetry \cite{top2, 10.1103/PhysRevB.83.125109}. While we restrict us, for the sake of clarity, to the SSH Hamiltonian (\ref{topologicalhamiltonian}), our emulation of a topological phase transition presented in the remainder also applies to such more general lattice Hamiltonians with two-dimensional unit cells. Similarly, extensions to two dimensional lattice models \cite{top2} are possible but require emulation of a higher-dimensional phase space. Finally, we expect that the emulation method is also adaptable to other symmetries and different choices of jump operators~\cite{10.1140/epjb/e2016-70325-x}.

The open system defined by (\ref{topologicalhamiltonian}) and (\ref{topologicaljumpoperator}) exhibits a dark state $|p\rangle\otimes|0\rangle$ [with $|0\rangle$ the ground state of the two-level system in Eq.~(\ref{topologicalhamiltonian})] when
\begin{eqnarray} \label{Eq:dark_state_condition}
h_x(p)=h_y(p)=0 ,
\end{eqnarray}
which is satisifed if $v=w$ and $p=\pi$. The jump-time evolution is in general not trace preserving when there are dark states, which renders the point (\ref{Eq:dark_state_condition}) singular in the Hamiltonian phase space. This then allows to classify lattice Hamiltonians (\ref{topologicalhamiltonian}) according to their winding about this singular point. Specifically, the Hamiltonian coefficients $h_x(p)$ and $h_{y}(p)$ in (\ref{topologicalhamiltonian}) describe a circle in the Hamiltonian phase space, which encircles the singular point when $v<w$, but does not when $v>w$, parting Hamiltonians into different topological classes depending on the relation of the hoppings, as shown in Fig.\ref{ssh}(b, c). The corresponding winding number can be defined as \cite{top2},
\begin{equation}\label{windingnumber}
W[\vec{h}(p)] = \oint \frac{\left[\frac{\partial h_x(p)}{\partial p}h_y(p)-\frac{\partial h_y(p)}{\partial p}h_x(p)\right]}{2\pi[h_x^2(p)+h_y^2(p)]}dp ,
\end{equation}
which counts the number of times the Hamiltonian encircles the singluar point $h_x(p)=h_y(p)=0$. While the winding number (\ref{windingnumber}) is applicable to any Hamiltonian $\vec{h}(p)$ with $h_z(p) \equiv 0$ (as well as to projections of general Hamiltonians $\vec{h}(p)$ into the $x-y$ plane), it evaluates for the SSH Hamiltonian~(\ref{topologicalhamiltonian}) either as 0 ($v>w$) or as 1 ($w>v$), depending on whether or not the singular point is encircled.

We now elaborate how the winding number defined in Eq.~(\ref{windingnumber}) can be reconstructed from the jump-time averaged density matrix. With the jump-time evolution relation in Eqs.~(\ref{jump-timeevolution}) and (\ref{defjumptimeevolutionoperator}), and exploiting that momentum is a good quantum number in our system, the jump-time propagator for the collective collapse jump operator $L_{\rm cc}$ simplifies to,
\begin{eqnarray}
\langle p|\rho_{n+1}|p'\rangle&=&K_{\rm cc}(p,p')\langle p|\rho_{n}|p'\rangle\nonumber\\
                              &=&[K_{\rm cc}(p,p')]^{n+1}\langle p|\rho_{0}|p'\rangle.
\end{eqnarray}
For out system described by Eqs.~(\ref{topologicalhamiltonian}) and (\ref{topologicaljumpoperator}), the jump-time propagator then evaluates as,
\begin{eqnarray} \label{Eq:jumptime_propagator}
K_{\rm cc}(p,p') &=&\left(\frac{\langle p|\otimes\langle 0  |\rho _{n}| 0\rangle\otimes |p'\rangle}{\langle p|\otimes\langle 0  |\rho _{0}| 0\rangle\otimes |p' \rangle}\right)^{\frac{1}{n}}\nonumber\\
&=&\frac{2[h_x(p)+ih_y(p)][h_x(p')-ih_y(p')]}{2[h^2(p)-h^2(p')]^2+[h^2(p)+h^{2}(p')]} , \nonumber\\
\end{eqnarray}
where the second line evaluates the propagator for the open system~(\ref{topologicalhamiltonian}) and (\ref{topologicaljumpoperator}), and $h^2(p)=h_x^2(p)+h_y^2(p)$. Note that the jump-time propagator~(\ref{Eq:jumptime_propagator}) is a scalar, as the jump time-averaged states necessarily reside on the sublattice denoted by $|0\rangle$.

It has been shown that the winding number $W[\vec{h}(p) ]$ can be recovered from the jump-time propagator~\cite{top2}, mediated through the jump-time phase
\begin{eqnarray}\label{k}
T_{\rm cc}[\vec{h}(p)]&=& \frac{i}{2\pi} \oint dp \left[ \frac{\partial K_{cc}(p,p')}{\partial p} \right]_{p'=p}.
\end{eqnarray}
Note that the integrand has a singular point at $(0,0)$, similar to the definition of the winding number (\ref{windingnumber}). Indeed, one finds that, with (\ref{Eq:jumptime_propagator}), the jump-time phase coincides with the winding number,
\begin{equation}\label{tk}
	T_{cc}[\vec{h}(p) ] = W[\vec{h}(p) ] .
\end{equation}

For actual lattice realizations of the open system~(\ref{topologicalhamiltonian}, \ref{topologicaljumpoperator}) the jump-time phase (\ref{k}) is observable in the average displacement $\langle \hat{x} \rangle_n$ of a single walker~\cite{top2}. Specifically, if the walker is initially located at lattice position $j_0$, then $\langle \hat{x} \rangle_n = a j_0 + a n T_{cc}[\vec{h}(p) ]$, where $a$ denotes the lattice constant. Here, instead, we aim at emulating the open system and its associated jumptime phase by a small system. To this end, we take advantage of the fact that, for the specific model under consideration, different momentum subspaces do not couple, and hence the effect of the jump-time propagator $K_{\rm cc}(p,p')$ can, for fixed $p$ and $p'$, be emulated with two coupled two-level systems (associated to $p$ and $p'$, respectively). Repeating this with a suitably chosen net of $p$ and $p'$ values then allows us to reconstruct the jumptime phase~(\ref{k}), and along with it, by virtue of~(\ref{tk}), the topological index of the underlying Hamiltonian. This will be elaborated in detail in Sec.~\ref{secnumsim}. We remark that this method to emulate topological phase-space properties has been successfully applied with fixed-time readout.~\cite{10.1016/j.scib.2021.02.035}. Here, we extend it to jump-time averaged quantum trajectories.

The emulation of the jump-time propagator through two two-level systems simplifies the experimental requirements significantly. In the following subsection, we will explain how the latter can be further lowered by virtue of a method to access jump-time averaged quantum states $\rho_n$ without actual monitoring.

\subsection{AUXILIARY SPACE}

While jump-time averaging $(\ref{Eq:jumptime-averaged_state})$ in principle requires monitoring, we can circumvent the continuous measurement along with the direct detection of quantum jumps if we complement the system by an auxiliary system that autonomously keeps track of the quantum jumps that occur on the system level. If the extended system is exposed to a suitably chosen environment (which does not have to emerge from monitoring), then, by construction, measuring the auxiliary system at fixed times allows us to infer the jump-time averaged system state of the target monitoring dynamics. In the following we elaborate this ancilla-enabled relation between jump-time averaged quantum states (of the target monitoring dynamics) and walltime-averaged quantum states (of the extended system subject to a Lindblad master equation), and demonstrate how it can be leveraged to recover the jumptime propagator $ $ of Eq.~(\ref{k}).

\begin{figure}[t]
\centering
\includegraphics[width=0.5\textwidth]{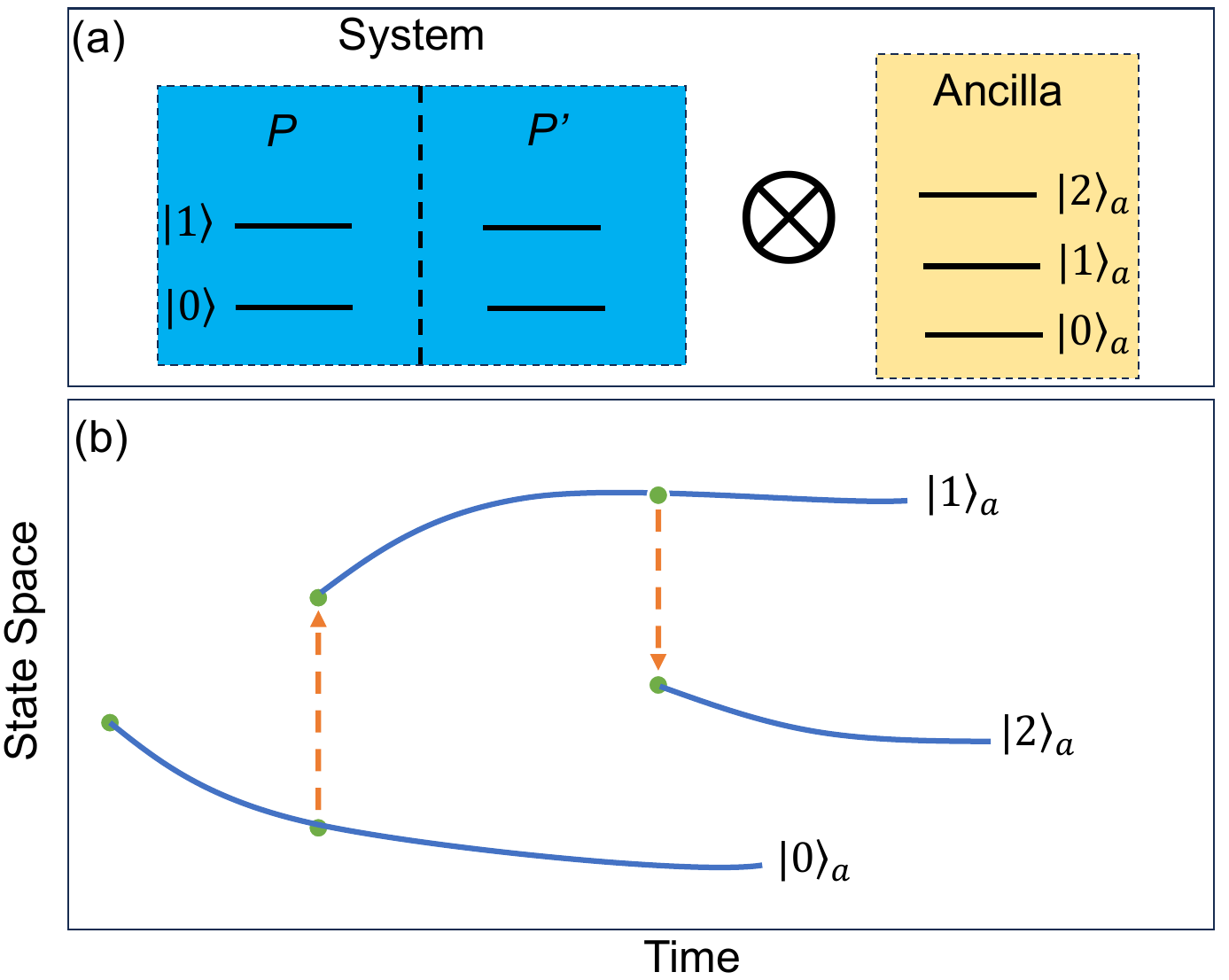} \caption{(a) Illustration of the combination of system and ancilla, which enables to emulate the topological phase transition observed in the dissipative Su-Schrieffer-Heeger model (see Fig.~\ref{ssh}). The system consists of a four-level system, which emulates two momentum subspaces of the lattice model, and a three-level auxiliary system, which records the quantum jump times in the system. (b) Different ancilla states collect quantum trajectories whose jump count matches the ancilla state label. The orange dashed lines depict the stochastically occurring quantum jumps.}
\label{f2}
\end{figure}

To establish the desired relation between jump-time averaged and walltime averaged quantum states, we complement the system by an auxiliary space, as illustrated in Fig.~\ref{f2}(a), where the Hilbert space dimension of the auxiliary space must accomodate the desired target jump count. The combination of system and auxiliary system then follows the Lindblad equation (\ref{Eq:Lindblad_master_equation}) with
\begin{equation}\label{ancillary}
\begin{split}
&\tilde{\rho}_{0} = \rho_{0}\otimes \left | 0 \right \rangle\left\langle0\right|_{\rm a}, \\
&\tilde{H} = H \otimes \mathbb{1}_a , \\
&\tilde{L}=L\otimes C_{+}, \\
\end{split}
\end{equation}
where $\rho_0$, $H$, and $L$ are the initial state, Hamiltonian, and jump operator of the original system, respectively, and $|0\rangle_{\rm a}$, $\mathbb{1}_a$, and $C_{+}$ are the ground state, identity operator, and raising operator of the auxiliary system. Note that we restrict us here to a single system jump operator, but the construction can easily be generalized to a family of system jump operators.

By construction, when a quantum jump occurs in the system, the Lindblad operator in Eq.~(\ref{ancillary}) also raises the state of the auxiliary system by one. A measurement in the canonical basis of the ancilla then projects onto the subset of (system-level) quantum trajectories that have, at the time of measurement, reached the jump count corresponding to the measured ancilla level, see Fig.~\ref{f2}(b). This way, the ancilla state can, if measured, give us access to system-level quantum trajectories with fixed jump count.

While this method can, with sufficiently large ancilla systems, give access to any finite jump count $n$, we restrict us here, for the sake of concreteness, to the case of $n=2$, as this is sufficient to recover the propagator $K_{\rm cc}(p,p')$ (recall that the latter, which itself is independent of $n$, mediates between two jump counts). This allows us to truncate the auxiliary system at the third level, 
\begin{equation}
\begin{split}
&C_{+}\left | n\right \rangle_{\rm a}=\left |n+1\right \rangle_{\rm a}, ~~n=0,1 , \\
&C_{+}\left | 2\right \rangle_{\rm a}=0 , \\
&\left \langle m| n \right \rangle_{\rm a} =\delta _{m,n},~~n,m=0,1,2.\\
\end{split}
\end{equation}
The lowest two auxiliary levels $|0\rangle_{\rm a}$ and $|1\rangle_{\rm a}$ record the states required to reconstruct the jumptime propagator, while the ancilla stays in the highest level $|2\rangle_{\rm a}$ after two jumps. Note that the ancilla state $|2\rangle_{\rm a}$ does not give access to the jump-time dynamics of the original system, but is still required to avoid detrimental backaction of the truncation on ancilla states $|0\rangle_{\rm a}$ and $|1\rangle_{\rm a}$(this will be elaborated below).

The jump-time averaged density matrices $\rho_{m}$ can now, by measuring the auxiliary space in the canonical basis, be obtained (see Appendix \ref{a2}) as follows:
\begin{equation}\label{m}
\begin{split}
&\rho _{m+1}=\gamma L\int_0^\infty dt \left \langle m |\tilde{\rho }_{t} | m\right \rangle_{\rm a} L^{\dag},~~m=0,1.\\
\end{split}
\end{equation}
We emphasize that $\tilde{\rho}_t$ is the walltime-averaged density matrix of the ancilla-extended system, which can be conveniently measured without the need to resort to monitoring, while $\rho_{n+1}$ is the desired jump-time averaged state of the target system. By comparing (\ref{jump-timeevolution}) and (\ref{m}), we find that the matrix element $\left \langle m |\tilde{\rho }_{t} | m\right \rangle_{\rm a}$ comprises the conditioned time evolution, induced by the effective Hamiltonian, from the time of the $m${\it th} jump to the readout time $t$. This shift reflects the fact that the $\rho_n$ in Eq.~(\ref{jump-timeevolution}) assumes direct access to the quantum jumps through detection, and hence knowledge of their time of occurrence, while the $\left \langle m |\tilde{\rho }_{t} | m\right \rangle_{\rm a}$ in Eq.~(\ref{m}) controls only the jump count but not the time lapse between jump and readout.

Instead of monitoring the system and counting the quantum jumps, the operational protocol to obtain $\rho_{n+1}$ via (\ref{m}) is: repeatedly run the experiment and measure the auxiliary system at a sufficiently dense grid of fixed times (one fixed time per run); at each time $t$ learn the quantum state components $\langle m |\tilde{\rho }_{t} | m \rangle$, $m=0,1$ (for instance, by quantum state tomography), whereas the ancilla outcome $|2\rangle_{\rm a}$ is discarded. Following (\ref{m}), the learned state components then allow us to approximate the associated jump-time averaged states, where the effect of the jump operators in (\ref{m}) is in general added by postprocessing. Note that the approximation becomes exact in the limit of an infinitely dense time grid. Moreover, for sufficiently large $t$, the norm of the state components $\left \langle m |\tilde{\rho }_{t} | m\right \rangle_{\rm a}$ generically becomes negligible, and we can safely truncate the upper integration limit.

From now on we focus on the open system (\ref{topologicalhamiltonian}, \ref{topologicaljumpoperator}), with the goal to reconstruct the jump-time propagator (\ref{Eq:jumptime_propagator}). Combining (\ref{Eq:jumptime_propagator}) with (\ref{m}), we can express the jump-time propagator as
\begin{eqnarray}\label{jtpropagatorwithax}
K_{\rm cc}(p, p')&=&\frac{\langle p|\otimes\langle 0  |\rho _{2}| 0\rangle\otimes |p'\rangle}{\langle p|\otimes\langle 0  |\rho _{1}| 0\rangle\otimes |p' \rangle}\nonumber\\
&=&\frac{\int_0^\infty dt\langle1|_{\rm a}\otimes\langle p|\otimes\langle0|L\tilde{\rho}_tL^{\dag}|0\rangle\otimes|p'\rangle\otimes|1\rangle_{\rm a}}{\int_0^\infty dt\langle0|_{\rm a}\otimes\langle p|\otimes\langle0|L\tilde{\rho}_tL^{\dag}|0\rangle\otimes|p'\rangle\otimes|0\rangle_{\rm a}}.\nonumber\\
\end{eqnarray}
According to Eqs.~(\ref{jtpropagatorwithax}) and (\ref{k}), we can emulate the topological phase transition with a four-level system and a three-level auxiliary system, as illustrated in Fig.~\ref{f2}(a). The three-level ancilla can separate trajectories with different jump times and give access to the density matrices in the jump-time domain, while the four-level system emulates the two two-level subspaces associated to different momenta $p$ and $p'$. Note that, to obtain the required element of the jump-time propagator in Eq.~(\ref{jtpropagatorwithax}), we only need to recover a specific element of the density matrix in the time domain, 
\begin{eqnarray}
&&\langle1|_{\rm a}\otimes\langle p|\otimes\langle0|L\tilde{\rho}_tL^{\dag}|0\rangle\otimes|p'\rangle\otimes|1\rangle_{\rm a}\nonumber\\
&=&\langle1|_{\rm a}\otimes\langle p|\otimes\langle1|\tilde{\rho}_t|1\rangle\otimes|p'\rangle\otimes|1\rangle_{\rm a},
\end{eqnarray}
which significantly reduces the reconstrucion cost at fixed times.

\section{Numerical verification of the proposal}\label{secnumsim}

\begin{figure}[t]
\centering
\includegraphics[width=0.38\textwidth]{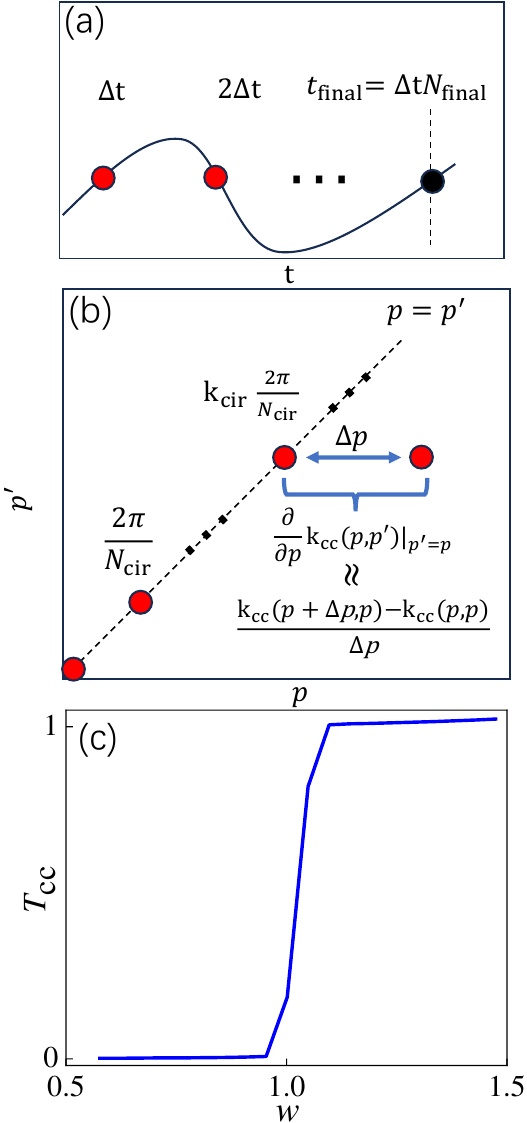} \caption{Approximate simulation of the jump-time topological phase transition. (a) Discretization and truncation in the time domain. The time integral in Eq.~(\ref{m}) is truncated at $t_{\rm final}$, resulting in a finite number of measurements $N_{\rm final}$. (b) Momentum space discretization. The integral of the momentum parameter $p$ in Eq.~(\ref{k}) is estimated with the sum of $N_{\rm cir}$ points in phase space. The differential of $p$ is substituted by a finite difference $\Delta p$. (c) The topological phase transition obtained with the parameters: $N_{\rm final}=300,a=1,v=1, N_{\rm cir}=500,\Delta p=0.01, t_{\rm final}=300$. We find that the topological phase transition is reliably recovered if the approximations are within an appropriate parameter range.}
\label{f3}
\end{figure}

We now numerically verify the validity of the proposal from the previous section. For any choice of $p$ and $p'$, the propagator $K_{\rm cc}(p, p')$ in Eq.~(\ref{jtpropagatorwithax}) only refers to the two respective momentum subspaces, which is why we can simulate the momentum degree of freedom in Eq.~(\ref{topologicalhamiltonian}) with a single two-level system, such that $|0\rangle\equiv|p\rangle$ and $|1\rangle\equiv|p'\rangle$. Accordingly, the momentum-dependent Hamiltonian is substituted by switchable pump terms. The total system is then described by the Hamiltonian,
\begin{eqnarray}\label{numH}
\tilde{H}&=&|0\rangle\langle 0|\otimes(h_{x}(p)\sigma_x+h_{y}(p)\sigma_y)\nonumber\otimes\mathbb{1}_{\rm a}\\
&&+|1\rangle\langle 1|\otimes(h_{x}(p')\sigma_x+h_{y}(p')\sigma_y)\otimes\mathbb{1}_{\rm a},
\end{eqnarray}
where the two projection operators $|1\rangle\langle1|$ and $|0\rangle\langle0|$ correspond to the two different momentum subspaces. The Hamiltonian coefficients $h_x(p)$ and $h_y(p)$ are as in Eq.~(\ref{topologicalhamiltonian}). The dissipation channel is described by the Lindblad operator
\begin{eqnarray}\label{numLindblad}
\tilde{L}&=&(|0\rangle\langle0|+|1\rangle\langle1|)\otimes\sigma_-\otimes C_+,
\end{eqnarray}
where $C_+$ is the raising operator of the ancillary space. As the jump-time propagator is related to the diagonal terms of the density matrix, we set the initial state to be a superposition state in the momentum subspace,
\begin{eqnarray}\label{numIS}
|\tilde{\psi}\rangle_0 =\frac{1}{\sqrt{2}} \left(|1\rangle + |0\rangle\right) \otimes |0\rangle\otimes|0\rangle_{\rm a} .
\end{eqnarray}
With Eqs.~(\ref{numH})-(\ref{numIS}), we can determine the density matrix $\tilde{\rho_t}$ of the total system. By virtue of Eq.~(\ref{m}), we can then approximate the jump-time averaged system states experimentally or numerically. To this end, we discretize the time integral according to
\begin{eqnarray}\label{apxjtdensitymatrix}
&\rho _{m+1}\approx\gamma L\sum_{k=1}^{N_{\rm final}}\Delta t\left \langle m |\tilde{\rho }_{k\Delta t} | m\right \rangle L^{\dag},~~m=0,1,
\end{eqnarray}
where we also truncate the upper integration limit at $t_{\rm final}=N_{\rm final}\Delta t$. In combination with the discretization $t=k\Delta t$, this then results in a finite number of measurement times, as illustrated in Fig.~\ref{f3} (a). Note that the probability for a trajectory to remain without quantum jump decreases exponentially with time. Therefore, we can omit the influence of the ``tail'' after the black dot in Fig.~\ref{f3} (a) for a large but finite $t_{\rm final}$.

Following a similar strategy, the topological order parameter in Eq.~(\ref{k}) is approximated as
\begin{eqnarray}\label{apxorderparameter}
T_{\rm cc}[\vec{h}(p)]&=&{\frac{i}{2\pi} \oint dp \left[ \frac{\partial K_{\rm cc}(p,p')}{\partial p} \right]_{p'=p}}\nonumber\\
                  &\approx&{\frac{i}{2\pi} \sum_{k_{\rm cir}=0}^{N_{\rm cir}}\frac{2\pi}{N_{\rm cir}} }\left[ \frac{\partial K_{cc}(p,p')}{\partial p} \right]_{p'=p=2\pi k_{\rm cir}/N_{\rm cir}} \nonumber\\
                  &\approx&\frac{i}{N_{\rm cir}}\sum_{k_{\rm cir}=0}^{N_{\rm cir}} \frac{K_{\rm cc}(\frac{2\pi k_{\rm cir}}{N_{\rm cir}}+\Delta p,\frac{2\pi k_{\rm cir}}{N_{\rm cir}})}{\Delta p} \nonumber\\ 
                  &&-\frac{i}{N_{\rm cir}}\sum_{k_{\rm cir}=0}^{N_{\rm cir}} \frac{K_{\rm cc}(\frac{2\pi k_{\rm cir}}{N_{\rm cir}},\frac{2\pi k_{\rm cir}}{N_{\rm cir}})}{\Delta p} , 
\end{eqnarray}
where we approximate the loop integral with the sum of $N_{\rm cir}$+1 points. Moreover, the derivative on $p$ is substituted by a finite difference $\Delta p$. The various discretizations are visualized in Figs.~\ref{f3}(a, b). The jumptime propagator at the probe points in (\ref{apxorderparameter}) is obtained from the approximate jumptime-average states (\ref{apxjtdensitymatrix}) by virtue of (\ref{Eq:jumptime_propagator}).

We simulate the approximated topological order parameter with Qutip~\cite{qut1,data} according to Eqs.~(\ref{numH})-(\ref{apxorderparameter}), and show the results in Fig.~\ref{f3}(c). Different values of the intercell hopping $w$ correspond to different loops in the phase space, as illustrated in Fig.~\ref{f2}, which follows from the structure of the original lattice. The numerical results indicate a jump of the order parameter at $w=1$, corresponding to $w=v$, which is consistent with the prediction from the previous section. This demonstrates the validity of our approximative proposal to emulate the jump-time topological phase transition with small systems. Note that all parameters are expressed in units of the decay rate $\gamma$. 

The experimental cost for measuring the phase transition can also be estimated with the discretized expressions in Eqs.~(\ref{apxjtdensitymatrix}) and (\ref{apxorderparameter}). According to Eqs.~(\ref{jtpropagatorwithax}), (\ref{apxjtdensitymatrix}), and (\ref{apxorderparameter}), estimating each $K_{\rm cc}$ requires the measurement of $2N_{\rm final}$ density matrix elements (or probabilities), each of which may require a sufficiently large number of measurements to suppress statistical fluctuations. As a result, the number of measurements for estimating $T_{\rm cc}$ is proportional to $2N_{\rm final}N_{\rm cir}$. In addition, estimating $T_{\rm cc}$ for different values of $w$ can further increase the  measurement cost. Therefore, it is important to identify the smallest $N_{\rm final}$ and $N_{\rm cir}$ required to observe a phase transition. To further clarify this question, we underpin the robustness of this topological phase transition under the approximations made here with more detailed studies in Appendix~\ref{approxcheck}.

\section{Experimental proposal for the simulation}\label{secexppro}

In this section, we briefly discuss the requirements for experimental realization of our proposal. The four-level system in Eq.~(\ref{ancillary}) can be simulated by two qubits with Ising-type coupling~\cite{10.1103/PhysRevB.75.140515,10.1103/PhysRevLett.127.050502}. One qubit simulates the two-level space, while the other qubit emulates the two momentum spaces. These two qubits need to feature an Ising-type coupling

\begin{eqnarray}
H_{\rm four}=\frac{\omega}{2}\sigma_z^{(p)}+\frac{\omega}{2}\sigma_z^{(\rm tl)}+E_{\rm couple}\sigma_z^{(\rm tl)}\sigma_z^{(p)} . 
\end{eqnarray}
The two eigenstates of the ``$p$'' qubit correspond to the two momentum subspaces associated with $p$ and $p'$, respectively, while the state within each momentum subspace is described by the ``tl'' qubit. Note that the effective frequency of the ``tl'' qubit is shifted from $(\omega+2E_{\rm couple})$ to $(\omega-2E_{\rm couple})$ by flipping the ``$p$'' qubit from $|1\rangle$ to $|0\rangle$. The momentum-dependent pumps in Eq.~(\ref{numH}) can be realized by choosing different pumping frequencies,
\begin{eqnarray}
H_{\rm fdp}&=&h_x(p)\sigma_x{\rm cos}([\omega+2E_{\rm couple}]t)\nonumber\\
&&+h_y(p)\sigma_y{\rm cos}([\omega+2E_{\rm couple}]t)\nonumber\\
&&+h_x(p')\sigma_x{\rm cos}([\omega-2E_{\rm couple}]t)\nonumber\\
&&+h_y(p')\sigma_y{\rm cos}([\omega-2E_{\rm couple}]t).
\end{eqnarray}
The required Lindblad operator in Eq.~(\ref{numLindblad}) can be realized by three-body coupling~\cite{10.1103/PhysRevLett.114.173602,10.1038/s41534-019-0219-y,10.1103/PhysRevLett.129.220501}, e.g.,
\begin{eqnarray}
H_{\rm thb}=g(\sigma_+^{(\rm tl)}\sigma^{\rm (a)}_- + \sigma_-^{(\rm tl)}\sigma^{\rm (a)}_+)(b^{\dag}+b),
\end{eqnarray}
with a strong loss on mode $b$ to prevent backward jumps. Note that the collective loss (\ref{numLindblad}) induced by the nonlinear coupling strength $H_{\rm thb}$ is usually one order of magnitude smaller than $g$. In the case of superconducting qubits, the experimentally achievable three-body nonlinear coupling is on the order of $10$MHz~\cite{10.1103/PhysRevLett.129.220501}, which limits the collective loss rate to $\gamma\approx1$MHz. To be able to clearly observe the jump-time topological phase transition, the rates of other, potentially detrimental noise sources thus need to be below the order of $100$kHz.

Regarding the three-level auxiliary system, we can introduce two auxiliary qubits and use the triplet states,
\begin{eqnarray}
\sigma_-^{({\rm a})}&=&\frac{1}{\sqrt{2}} (\sigma_-^{({\rm a},1)}+\sigma_-^{({\rm a},2)}),\nonumber\\
|0\rangle&=&|0\rangle\otimes|0\rangle,\nonumber\\
|1\rangle&=&\frac{1}{\sqrt{2}}(|1\rangle\otimes|0\rangle+|0\rangle\otimes|1\rangle),\nonumber\\
|2\rangle&=&|1\rangle\otimes|1\rangle.
\end{eqnarray}

\begin{figure}[t]
\centering
\includegraphics[width=0.48\textwidth]{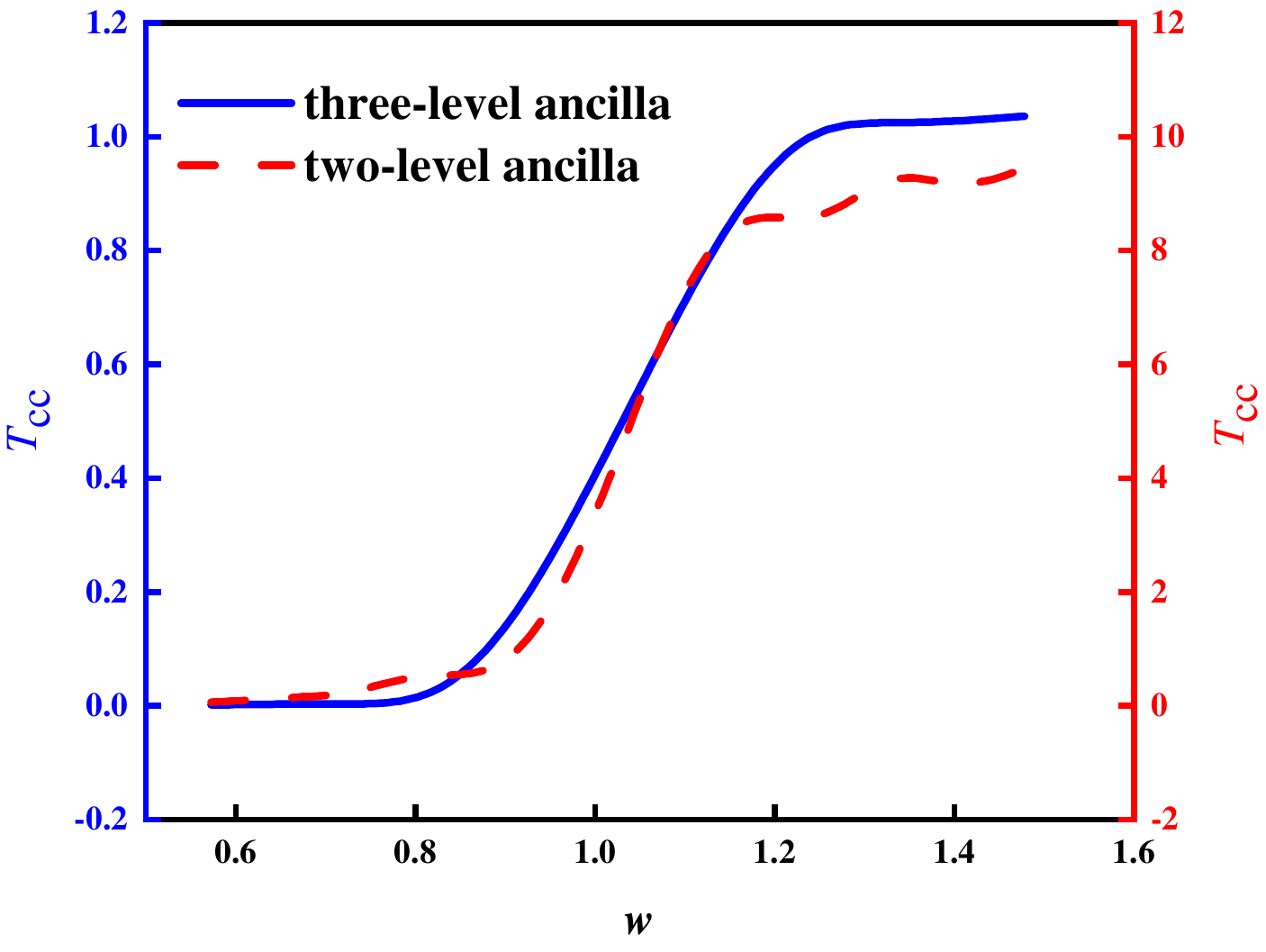} \caption{Role of the third ancilla level. Shown is the topological phase transition simulated with a three-level ancilla (blue curve and left vertical axis) and a two-level ancilla (red curve and right vertical axis). The simluation parameters are: $N_{\rm final}=20$, $N_{\rm cir}=20$, and $\Delta p=0.01$. One finds that a simulation with a smaller auxiliary system still remains sensitive to the topological phase transition, but there is a normalization problem.}
\label{f6}
\end{figure}
This could be further simplified by scaling down the ancilla to a two-level system, so that the number of qubits is effectively reduced from four to three. To assess this simplification, we substitute the auxiliary system with a qubit and numerically check the quality of the retrieval of the topological order parameter $T_{\rm cc}$. As illustrated in Fig.~\ref{f6}, the $T_{cc}$ curve follows the correct pattern but takes much larger values without the third level. The reason for the inflated values of $T_{\rm cc}$ is that the normalization of the jump-time averaged density matrix $\rho_{n}$ is related to the average waiting period before a jump. These waiting periods, however, become infinite in the absence of the third level.  

\section{Conclusions}
\label{conclusion}

We elaborated a method to simulate the topological phase transition in the jump-time domain with a small system. The method maps different momentum components of the original translation-invariant lattice model onto two separate but coupled two level systems, which drastically reduces the complexity of the required experimental setup. Moreover, by introducing an additional three-dimensional auxiliary space, access to jump-time averaged system states is obtained through measuring the auxiliary system at fixed times, which allows us to circumvent the need to implement weak continuous measurements.

In contrast to the original lattice model, where the topological phase is observed in the transport behavior of the monitored system, our method directly emulates the jump-time phase, which coincides with the winding number as the topological index that characterizes the open system. This emulation relies on scanning through a sufficiently fine-grained parameter grid, both in momentum and in time, where, for each parameter setting, the associated quantum state is reconstructed through quantum state tomography.

Due to the comparative simplicity of the proposed simulation, realizations are conceivable on a wide array of experimental platforms. In particular, with superconducting qubits the required Ising-type and three-body interactions would be readily available~\cite{10.1103/PhysRevB.75.140515,10.1103/PhysRevLett.127.050502,10.1103/PhysRevLett.114.173602,10.1038/s41534-019-0219-y,10.1103/PhysRevLett.129.220501}. However, other platforms, such as optomechanical systems~\cite{10.1038/nature10261}, quantum dots~\cite{10.1063/1.102525,10.1109/3.910456,10.1016/j.physrep.2004.01.004}, or hybrid systems~\cite{PhysRevLett.130.073602}, may also provide viable implementation paths.

{While we demonstrate our simulation method with a one-dimensional SSH lattice model,  the method can be adapted to other lattice models and higher dimensions. Moreover, simulations of arbitrary jump counts are possible by increasing the dimension of the auxiliary system accordingly. Finally, let us point out that the established relationship~(\ref{m}) between jump-time averaged states of monitored systems and wall-time averaged states of (unmonitored) ancilla-extended systems is neither restricted to lattice systems, nor to the emulation method presented here, but may also prove useful in other studies of the jump-time dynamics and its consequences.

\begin{acknowledgments}
This research is supported by Zhejiang Provincial Natural Science Foundation of China under Grant No.~LQ24A050002; the National Natural Science Foundation of China (NSFC) (Grant No.~12405028); the Innovation Program for Quantum Science and Technology (2023ZD0300904). C.G. is partially supported by RIKEN Incentive Research Projects. A.X.C. is partially supported by the National Natural Science Foundation of China (Grants No.12175199), and Foundation of Department of Science and Technology of Zhejiang Province (Grant No. 2022R52047). 
\end{acknowledgments}

\appendix
\section{Trajectory decomposition of $\rho_{t}$}
\label{a1}

We here show that the decomposition~(\ref{Eq:formal_Lindblad_solution}) of the density matrix into quantum trajectories with different jump times is a solution of the master equation~(\ref{Eq:Lindblad_master_equation}).Let us consider the formal decomposition~(\ref{Eq:formal_Lindblad_solution}),
\begin{equation}
\begin{split}
\rho _{t}=&\sum_{n=0}^{\infty} \int_{0}^{t} dt_{n}\int_{0}^{t_{n}} dt_{n-1}\dots \int_{0}^{t_{2}} dt_{1}\sum _{j_{1}\dots j_{n}\in I}\rho^{t} _{j_{n}\dots j_{1}}(\left\{t_{i} \right\}).\\
\end{split}
\end{equation}
We now take the time derivative,
\begin{equation}
\begin{split}
\dot\rho _{t}=&\sum_{n=1}^{\infty} \int_{0}^{t} dt_{n-1}\dots \int_{0}^{t_{2}} dt_{1}\sum _{j_{1}\dots j_{n}\in I}\rho^{t} _{j_{n}\dots j_{1}}(\left\{t_{i} \right\})|_{t_{n}=t}\\
&+\sum_{n=0}^{\infty}\int_{0}^{t} dt_{n}\int_{0}^{t_{n}} dt_{n-1}\dots \int_{0}^{t_{2}} dt_{1}\\
&\hspace{2mm} \sum _{j_{1}\dots j_{n}\in I} \big((-i H_{\rm eff}/\hbar) \rho^{t} _{j_{n}\dots j_{1}}(\left\{t_{i} \right\})\\
&\hspace{15mm}+\rho^{t} _{j_{n}\dots j_{1}}(\left\{t_{i} \right\})iH_{\rm eff}^\dagger/\hbar \big).
\end{split}
\end{equation}
If we mark the first line and the second line as $a$ and $b$, respectively, the second line reads
\begin{equation}
\begin{split}
&b=(-iH_{\rm eff}/\hbar)\rho_{t}+\rho_{t}(iH_{\rm eff}^\dagger/\hbar),
\end{split}
\end{equation}
while the first line reads
\begin{equation}
\begin{split}
a&=\sum_{n=1}^{\infty} \int_{0}^{t} dt_{n-1}\dots \int_{0}^{t_{2}} dt_{1}\sum _{j_{1}\dots j_{n}\in I}\rho^{t} _{j_{n}\dots j_{1}}(\left\{t_{i} \right\})|_{t_{n}=t}\\
&=\sum_{n=1}^{\infty} \int_{0}^{t} dt_{n-1}\dots \int_{0}^{t_{2}} dt_{1}\sum _{j_{1}\dots j_{n}\in I}\mathcal{J}_{j_{n}}\rho^{t} _{j_{n-1}\dots j_{1}}(\left\{t_{i} \right\})|_{t_{n}=t}\\
&=\sum_{J_{n}} J_{J_{n}}\rho _{t}.\\
\end{split}
\end{equation}
We thus can write
\begin{equation}
\begin{split}
&\dot\rho _{t}=\sum_{J_{n}} J_{J_{n}}\rho _{t}+(-iH_{\rm eff}/\hbar)\rho_{t}+\rho_{t}(iH_{\rm eff}^\dagger/\hbar).\\
\end{split}
\end{equation}
With the (non-Hermitian) effective Hamiltonian \cite{tra3}
\begin{equation}
H_{\rm eff} =H-i\hbar \frac{\gamma}{2} \sum_{j\in I}L_{j}^{\dagger} L_{j},
\end{equation}
it follows that
\begin{align}
\dot\rho _{t}=&\sum_{J_{n}} J_{J_{n}}\rho _{t}+(-iH/\hbar-\frac{\gamma}{2} \sum_{j\in I}L_{j}^{\dagger } L_{j})\rho _{t}\nonumber \\
&+\rho _{t}(iH/\hbar-\frac{\gamma}{2}\sum_{j\in I}L_{j}^{\dagger } L_{j})\nonumber \\
=&-(i/\hbar) (H\rho _{t}-\rho _{t}H)\nonumber \\
&+\gamma \sum_{j\in I} \big(L_{j}\rho L_{j}^{\dagger }-1/2(L_{j}^{\dagger }L_{j}\rho_{t}+ \rho L_{j}^{\dagger }L_{j})\big),\nonumber \\
\end{align}
that is, $\rho(t)$ as in~(\ref{Eq:formal_Lindblad_solution}) solves the master equation~(\ref{Eq:Lindblad_master_equation}). This concludes our proof that the density matrix $\rho_t$ can be decomposed into trajectories with different jump times.

\section{Obtaining $\rho_{m}$ from $\tilde{\rho}_t$}
\label{a2}
Here, we show how to obtain the density matrix in the jump-time domain by fixed-time measurements on the auxiliary system.

Consider the following projective measurement on the auxiliary space,
\begin{eqnarray}
&&\langle m  |\tilde{\rho }_{t} | m \rangle_{\rm a} \nonumber\\
&=&\langle m  |\sum_{n=0}^{\infty} \int_{0}^{t} dt_{n}(\prod_{i=1}^{n-1}\int_{0}^{t_{i+1}} dt_{i})\sum _{j_{1}\dots j_{n}\in I}\tilde{\rho}^{t} _{j_{n}\dots j_{1}}(\left\{t_{i} \right\})| m \rangle_{\rm a} ,\nonumber\\
&=&\sum_{n=0}^{\infty} \int_{0}^{t} dt_{n}\int_{0}^{t_{n}} dt_{n-1}\dots \int_{0}^{t_{2}} dt_{1} a_n \langle m| C_+^n |0 \rangle_{\rm a}\langle 0|C_-^n |m \rangle_{\rm a} \nonumber\\
&=&\int_{0}^{t} dt_{m}\int_{0}^{t_{n}} dt_{n-1}\dots \int_{0}^{t_{2}} dt_{1}a_m,\nonumber\\
\end{eqnarray}
where
\begin{eqnarray}
a_n&=&\sum _{j_{1}\dots j_{n}\in I} \mathcal{U}_{t-t_{n}}\mathcal{J}_{j_{n}}\mathcal{U}_{t_{n}-t_{n-1}}\mathcal{J}_{j_{n-1}}\dots \mathcal{J}_{j_{1}}\mathcal{U}_{t_{1}} \rho_{0}.\nonumber\\
\end{eqnarray}
Here, $|m\rangle_{\rm a}$ represent the eigenstates of the auxiliary space, with $m=0, 1, 2$. On the other hand, from (\ref{Eq:jumptime-averaged_state}) we can derive the relation
\begin{eqnarray}
&&\int_{0}^{\infty}dt\sum_{k}\mathcal{J}_k\int_{0}^{t} dt_{m}\int_{0}^{t_{n}} dt_{n-1}\dots \int_{0}^{t_{2}} dt_{1}a_m=\rho_{m+1}.\nonumber\\
\end{eqnarray}
Combining these pieces, we obtain the desired relation (\ref{m}) between the jump-time averaged and wall-time averaged quantum states,
\begin{eqnarray}
\rho_{m}=\int_0^{\infty}\sum_k \mathcal{J}_k\left \langle m-1  |\tilde{\rho }_{t} | m-1\right \rangle.
\end{eqnarray}
This relation shows how the density matrix in the jump-time domain $\rho_m$ can be obtained from projective measurements on the auxiliary space $\langle m-1  |\tilde{\rho }_{t} | m-1 \rangle$.

\section{Influence of the approximations}\label{approxcheck}

The conditions for the validity of the approximations in Eqs.~(\ref{apxjtdensitymatrix}) and (\ref{apxorderparameter}) are closely related to the experimental feasibility of this proposal. Therefore, we here numerically explore the validity limits of these approximations and provide the parameter range in which the topological phase transition can be reliably measured.

\begin{figure}[t]
\centering
\includegraphics[width=0.45\textwidth]{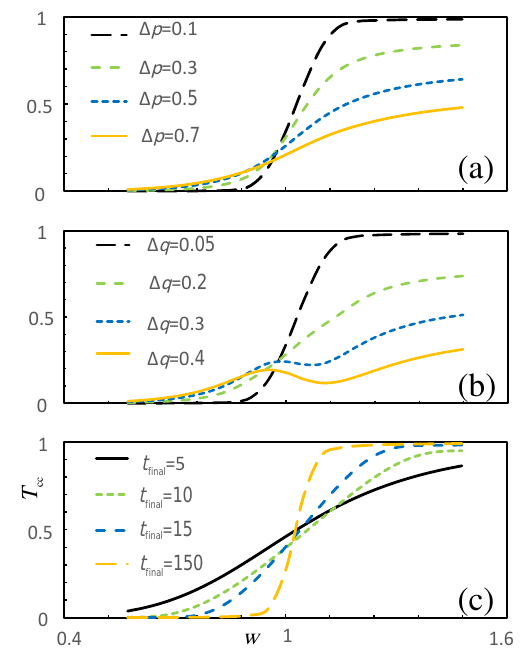} \caption{(a) The influence of approximating the differential with a finite difference $\Delta p$ is studied with the parameters: $N_{\rm final}=200, a=1, v=1, N_{\rm cir}=50,\Delta q=0.02,t_{\rm final}=50$. The larger $\Delta p$, the smoother the line is, the smaller $\Delta p$, the steeper the line is, and when $\Delta p<0.3$, the line jumps significantly around the phase transition at $w=1$. (b) The influence of the approximation $({\rm lim}_{q\rightarrow p}q\approx p+\Delta q)$ is studied with the parameters: $N_{\rm final}=200,a=1,v=1, N_{\rm cir}=50,\Delta p=0.01,t_{\rm final}=50$, when $\Delta q>0.3$. Similar to $\Delta p$,$\Delta q$ is related to the sharpness of the transition. In addition, the curve develops spurious oscillations when  $\Delta q>0.2$. (c) The results for finite sampling time $t_{\rm final}$ are shown. The other parameters are as follows: $N_{\rm final}=200,a=1,v=1,,N_{\rm cir}=50,\Delta p=0.01,\Delta q=0.02$. As $t_{\rm final}$ gets larger, the transition gets sharper}.
\label{f4}
\end{figure}

We first consider the role of the finite changes of the momentum parameters $\Delta p$ and a finite accuracy parameter error $\Delta q\equiv p'-p$, which both should be infinitely small in the ideal limit. These two parameters describe the precision necessary for adjusting the pumps in the system. As illustrated in Figs.~\ref{f4} (a) and (b), both parameters are related to the ``height'' of the jump. For $\Delta p \lesssim 0.3$, there is a significant jump, while for larger $\Delta p$, this jump gradually disappears. From Fig.~\ref{f4}(b), we can find that the results are more sensitive to $\Delta q$. For $\Delta q \approx 0.2$, the jump is still not obvious. In addition, the curves start to oscillate for $\Delta q=0.3$ and $\Delta q=0.4$. Therefore, the precision required to observe the phase transition is $\Delta p \lesssim 0.1$ and $\Delta q \lesssim 0.05$, respectively. As $\Delta p$ and $\Delta q$ correspond to the same quantity in the system, the parameter $p$ of the system~(\ref{numH}) should be controlled at a precision $p \lesssim 0.05$.

After the momentum parameters, we now consider the influence of the finite obervation time $t_{\rm final}$ in Fig.~\ref{f4}(c), which ideally should be infinite to account for the influence of quantum jumps at any time points. One finds that the parameter $t_{\rm final}$ mainly affects the steepness of the jump of the order parameter $T_{\rm cc}$. When $t_{\rm final} \gtrsim 15~({\rm in\ units\ of\ }1/\gamma)$, the results exhibit a significant jump. For $t_{\rm final}=150$, no visible improvements can be found compared to $t_{\rm final}=50$ in Figs.~\ref{f4}(a) and (b). Therefore, $t_{\rm final} \gtrsim 50$ should be sufficiently long for observing the topological phase transition.

\begin{figure}[t]
\centering
\includegraphics[width=0.4\textwidth]{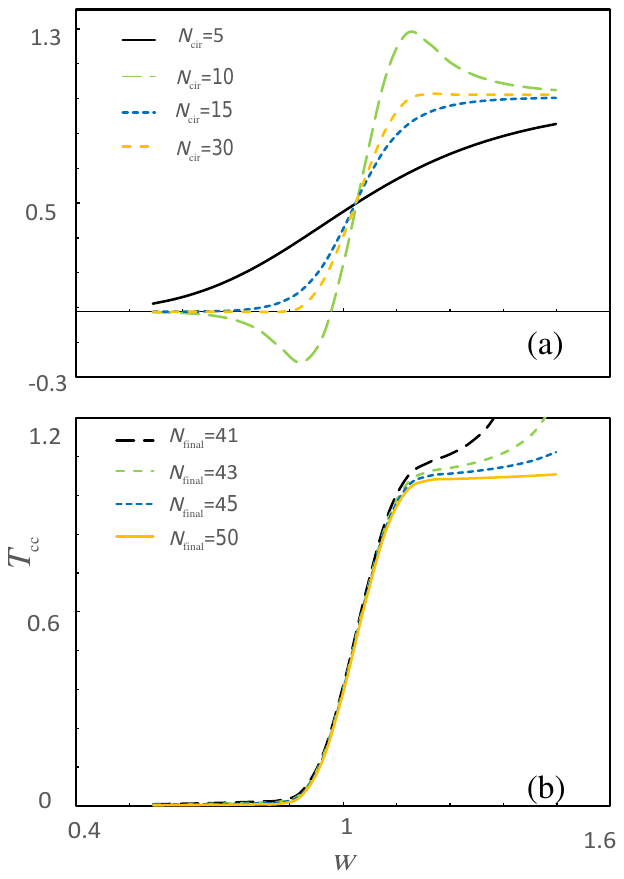} \caption{Estimation of the minimum number of required measurement points in the time domain and in the momentum domain. (a) Results with different numbers of measurement points in the momentum domain are compared with the parameters: $N_{\rm final}=200, a=1, v=1, \Delta p=0.01, \Delta q=0.02, t_{\rm final}=50$. When $N_{\rm cir}$ is close to 10, the curve significantly diverges from the ideal case, while $N_{\rm cir}=30$ can result in a comparatively clear phase transition. (b) The influence of the number of measurements in the time domain is studied with the parameters: $a=1, v=1, N_{\rm cir}=50,\Delta p=0.01,\Delta q=0.02,t_{\rm final}=50$. If $N_{\rm final} < 50$, the curve will keep increasing instead of converging to 1. $N_t$ mainly controls the asymptotic properties of the curve.}
\label{f5}
\end{figure}

Finally, we consider the role of the number of measurements in the time domain $N_{\rm final}$ and the one in the phase space $N_{\rm circle}$. These two parameters can increase the required number of measurements and resolutions in the corresponding domains. In Fig.~\ref{f5}(a), we show the estimated order parameters $T_{\rm cc}$ with different measurement numbers in the momentum domain $p$. We find that, for $N_{\rm cir}=15$, the results exhibit the desired jump. When the numbers are smaller, the curves significantly deviate from an ideal phase transition. Specifically, for even $N_{\rm cir}$ the values are larger than the ideal one for $w>1$ (phase transition point) and smaller than the ideal one for $w<1$. On the other hand, odd $N_{\rm cir}$ show opposite deviations. Note that $N_{\rm cir}$ corresponds to the resolution in momentum space, $\Delta p=2\pi/N_{\rm cir}$, but this required resolution is low compared to the one found in Figs.~\ref{f4}(a) and (b).

The measurement number in the time domain $N_{\rm final}$ mainly affects the results with $w>1$, so that it is less relevant to the detection of the phase transition point. However, the number required for obtaining a shape close to the ideal one is larger compared to the measurement number $N_{\rm cir}$.

In this Appendix, we numerically estimated the parameters required for simulating the topological phase transition with a small system. Specifically, the required control precision in the effective momentum space is $\Delta p \lesssim 0.05$; the minimal time period to monitor the system is $t_{\rm final} \gtrsim 150$; the necessary measurement numbers in the time domain and the effective momentum domain are $N_{\rm final}>40$ and $N_{\rm cir}>15$, respectively. 



\bibliography{reference}

\begin{thebibliography}{84}%
\makeatletter
\providecommand \@ifxundefined [1]{%
 \@ifx{#1\undefined}
}%
\providecommand \@ifnum [1]{%
 \ifnum #1\expandafter \@firstoftwo
 \else \expandafter \@secondoftwo
 \fi
}%
\providecommand \@ifx [1]{%
 \ifx #1\expandafter \@firstoftwo
 \else \expandafter \@secondoftwo
 \fi
}%
\providecommand \natexlab [1]{#1}%
\providecommand \enquote  [1]{``#1''}%
\providecommand \bibnamefont  [1]{#1}%
\providecommand \bibfnamefont [1]{#1}%
\providecommand \citenamefont [1]{#1}%
\providecommand \href@noop [0]{\@secondoftwo}%
\providecommand \href [0]{\begingroup \@sanitize@url \@href}%
\providecommand \@href[1]{\@@startlink{#1}\@@href}%
\providecommand \@@href[1]{\endgroup#1\@@endlink}%
\providecommand \@sanitize@url [0]{\catcode `\\12\catcode `\$12\catcode `\&12\catcode `\#12\catcode `\^12\catcode `\_12\catcode `\%12\relax}%
\providecommand \@@startlink[1]{}%
\providecommand \@@endlink[0]{}%
\providecommand \url  [0]{\begingroup\@sanitize@url \@url }%
\providecommand \@url [1]{\endgroup\@href {#1}{\urlprefix }}%
\providecommand \urlprefix  [0]{URL }%
\providecommand \Eprint [0]{\href }%
\providecommand \doibase [0]{https://doi.org/}%
\providecommand \selectlanguage [0]{\@gobble}%
\providecommand \bibinfo  [0]{\@secondoftwo}%
\providecommand \bibfield  [0]{\@secondoftwo}%
\providecommand \translation [1]{[#1]}%
\providecommand \BibitemOpen [0]{}%
\providecommand \bibitemStop [0]{}%
\providecommand \bibitemNoStop [0]{.\EOS\space}%
\providecommand \EOS [0]{\spacefactor3000\relax}%
\providecommand \BibitemShut  [1]{\csname bibitem#1\endcsname}%
\let\auto@bib@innerbib\@empty
\bibitem [{\citenamefont {Hasan}\ and\ \citenamefont {Kane}(2010)}]{RevModPhys.82.3045}%
  \BibitemOpen
  \bibfield  {author} {\bibinfo {author} {\bibfnamefont {M.~Z.}\ \bibnamefont {Hasan}}\ and\ \bibinfo {author} {\bibfnamefont {C.~L.}\ \bibnamefont {Kane}},\ }\bibfield  {title} {\bibinfo {title} {{\it Colloquium}: {Topological} insulators},\ }\href {https://doi.org/10.1103/RevModPhys.82.3045} {\bibfield  {journal} {\bibinfo  {journal} {Rev. Mod. Phys.}\ }\textbf {\bibinfo {volume} {82}},\ \bibinfo {pages} {3045} (\bibinfo {year} {2010})}\BibitemShut {NoStop}%
\bibitem [{\citenamefont {Qi}\ and\ \citenamefont {Zhang}(2011)}]{RevModPhys.83.1057}%
  \BibitemOpen
  \bibfield  {author} {\bibinfo {author} {\bibfnamefont {X.-L.}\ \bibnamefont {Qi}}\ and\ \bibinfo {author} {\bibfnamefont {S.-C.}\ \bibnamefont {Zhang}},\ }\bibfield  {title} {\bibinfo {title} {Topological insulators and superconductors},\ }\href {https://doi.org/10.1103/RevModPhys.83.1057} {\bibfield  {journal} {\bibinfo  {journal} {Rev. Mod. Phys.}\ }\textbf {\bibinfo {volume} {83}},\ \bibinfo {pages} {1057} (\bibinfo {year} {2011})}\BibitemShut {NoStop}%
\bibitem [{\citenamefont {Ryu}\ \emph {et~al.}(2010)\citenamefont {Ryu}, \citenamefont {Schnyder}, \citenamefont {Furusaki},\ and\ \citenamefont {Ludwig}}]{Ryu_2010}%
  \BibitemOpen
  \bibfield  {author} {\bibinfo {author} {\bibfnamefont {S.}~\bibnamefont {Ryu}}, \bibinfo {author} {\bibfnamefont {A.~P.}\ \bibnamefont {Schnyder}}, \bibinfo {author} {\bibfnamefont {A.}~\bibnamefont {Furusaki}},\ and\ \bibinfo {author} {\bibfnamefont {A.~W.~W.}\ \bibnamefont {Ludwig}},\ }\bibfield  {title} {\bibinfo {title} {Topological insulators and superconductors: tenfold way and dimensional hierarchy},\ }\href {https://doi.org/10.1088/1367-2630/12/6/065010} {\bibfield  {journal} {\bibinfo  {journal} {New Journal of Physics}\ }\textbf {\bibinfo {volume} {12}},\ \bibinfo {pages} {065010} (\bibinfo {year} {2010})}\BibitemShut {NoStop}%
\bibitem [{\citenamefont {Asb{\'o}th}\ \emph {et~al.}(2016)\citenamefont {Asb{\'o}th}, \citenamefont {Oroszl{\'a}ny},\ and\ \citenamefont {P{\'a}lyi}}]{asboth2016short}%
  \BibitemOpen
  \bibfield  {author} {\bibinfo {author} {\bibfnamefont {J.~K.}\ \bibnamefont {Asb{\'o}th}}, \bibinfo {author} {\bibfnamefont {L.}~\bibnamefont {Oroszl{\'a}ny}},\ and\ \bibinfo {author} {\bibfnamefont {A.}~\bibnamefont {P{\'a}lyi}},\ }\href {https://doi.org/10.1007/978-3-319-25607-8} {\emph {\bibinfo {title} {A short course on topological insulators:Band Structure and Edge States in One and Two Dimensions}}}\ (\bibinfo  {publisher} {Springer},\ \bibinfo {year} {2016})\BibitemShut {NoStop}%
\bibitem [{\citenamefont {Bardyn}\ \emph {et~al.}(2013)\citenamefont {Bardyn}, \citenamefont {Baranov}, \citenamefont {Kraus}, \citenamefont {Rico}, \citenamefont {İmamoğlu}, \citenamefont {Zoller},\ and\ \citenamefont {Diehl}}]{Bardyn_2013}%
  \BibitemOpen
  \bibfield  {author} {\bibinfo {author} {\bibfnamefont {C.-E.}\ \bibnamefont {Bardyn}}, \bibinfo {author} {\bibfnamefont {M.~A.}\ \bibnamefont {Baranov}}, \bibinfo {author} {\bibfnamefont {C.~V.}\ \bibnamefont {Kraus}}, \bibinfo {author} {\bibfnamefont {E.}~\bibnamefont {Rico}}, \bibinfo {author} {\bibfnamefont {A.}~\bibnamefont {İmamoğlu}}, \bibinfo {author} {\bibfnamefont {P.}~\bibnamefont {Zoller}},\ and\ \bibinfo {author} {\bibfnamefont {S.}~\bibnamefont {Diehl}},\ }\bibfield  {title} {\bibinfo {title} {Topology by dissipation},\ }\href {https://doi.org/10.1088/1367-2630/15/8/085001} {\bibfield  {journal} {\bibinfo  {journal} {New Journal of Physics}\ }\textbf {\bibinfo {volume} {15}},\ \bibinfo {pages} {085001} (\bibinfo {year} {2013})}\BibitemShut {NoStop}%
\bibitem [{\citenamefont {Bardyn}\ \emph {et~al.}(2018)\citenamefont {Bardyn}, \citenamefont {Wawer}, \citenamefont {Altland}, \citenamefont {Fleischhauer},\ and\ \citenamefont {Diehl}}]{PhysRevX.8.011035}%
  \BibitemOpen
  \bibfield  {author} {\bibinfo {author} {\bibfnamefont {C.-E.}\ \bibnamefont {Bardyn}}, \bibinfo {author} {\bibfnamefont {L.}~\bibnamefont {Wawer}}, \bibinfo {author} {\bibfnamefont {A.}~\bibnamefont {Altland}}, \bibinfo {author} {\bibfnamefont {M.}~\bibnamefont {Fleischhauer}},\ and\ \bibinfo {author} {\bibfnamefont {S.}~\bibnamefont {Diehl}},\ }\bibfield  {title} {\bibinfo {title} {Probing the {Topology} of {Density Matrices}},\ }\href {https://doi.org/10.1103/PhysRevX.8.011035} {\bibfield  {journal} {\bibinfo  {journal} {Phys. Rev. X}\ }\textbf {\bibinfo {volume} {8}},\ \bibinfo {pages} {011035} (\bibinfo {year} {2018})}\BibitemShut {NoStop}%
\bibitem [{\citenamefont {Lieu}\ \emph {et~al.}(2020)\citenamefont {Lieu}, \citenamefont {McGinley},\ and\ \citenamefont {Cooper}}]{PhysRevLett.124.040401}%
  \BibitemOpen
  \bibfield  {author} {\bibinfo {author} {\bibfnamefont {S.}~\bibnamefont {Lieu}}, \bibinfo {author} {\bibfnamefont {M.}~\bibnamefont {McGinley}},\ and\ \bibinfo {author} {\bibfnamefont {N.~R.}\ \bibnamefont {Cooper}},\ }\bibfield  {title} {\bibinfo {title} {{Tenfold Way} for {Quadratic Lindbladians}},\ }\href {https://doi.org/10.1103/PhysRevLett.124.040401} {\bibfield  {journal} {\bibinfo  {journal} {Phys. Rev. Lett.}\ }\textbf {\bibinfo {volume} {124}},\ \bibinfo {pages} {040401} (\bibinfo {year} {2020})}\BibitemShut {NoStop}%
\bibitem [{\citenamefont {Dangel}\ \emph {et~al.}(2018)\citenamefont {Dangel}, \citenamefont {Wagner}, \citenamefont {Cartarius}, \citenamefont {Main},\ and\ \citenamefont {Wunner}}]{PhysRevA.98.013628}%
  \BibitemOpen
  \bibfield  {author} {\bibinfo {author} {\bibfnamefont {F.}~\bibnamefont {Dangel}}, \bibinfo {author} {\bibfnamefont {M.}~\bibnamefont {Wagner}}, \bibinfo {author} {\bibfnamefont {H.}~\bibnamefont {Cartarius}}, \bibinfo {author} {\bibfnamefont {J.}~\bibnamefont {Main}},\ and\ \bibinfo {author} {\bibfnamefont {G.}~\bibnamefont {Wunner}},\ }\bibfield  {title} {\bibinfo {title} {Topological invariants in dissipative extensions of the {Su-Schrieffer-Heeger} model},\ }\href {https://doi.org/10.1103/PhysRevA.98.013628} {\bibfield  {journal} {\bibinfo  {journal} {Phys. Rev. A}\ }\textbf {\bibinfo {volume} {98}},\ \bibinfo {pages} {013628} (\bibinfo {year} {2018})}\BibitemShut {NoStop}%
\bibitem [{\citenamefont {Minganti}\ \emph {et~al.}(2019)\citenamefont {Minganti}, \citenamefont {Miranowicz}, \citenamefont {Chhajlany},\ and\ \citenamefont {Nori}}]{PhysRevA.100.062131}%
  \BibitemOpen
  \bibfield  {author} {\bibinfo {author} {\bibfnamefont {F.}~\bibnamefont {Minganti}}, \bibinfo {author} {\bibfnamefont {A.}~\bibnamefont {Miranowicz}}, \bibinfo {author} {\bibfnamefont {R.~W.}\ \bibnamefont {Chhajlany}},\ and\ \bibinfo {author} {\bibfnamefont {F.}~\bibnamefont {Nori}},\ }\bibfield  {title} {\bibinfo {title} {Quantum exceptional points of non-{Hermitian Hamiltonians} and {Liouvillians}: The effects of quantum jumps},\ }\href {https://doi.org/10.1103/PhysRevA.100.062131} {\bibfield  {journal} {\bibinfo  {journal} {Phys. Rev. A}\ }\textbf {\bibinfo {volume} {100}},\ \bibinfo {pages} {062131} (\bibinfo {year} {2019})}\BibitemShut {NoStop}%
\bibitem [{\citenamefont {Song}\ \emph {et~al.}(2019)\citenamefont {Song}, \citenamefont {Yao},\ and\ \citenamefont {Wang}}]{PhysRevLett.123.170401}%
  \BibitemOpen
  \bibfield  {author} {\bibinfo {author} {\bibfnamefont {F.}~\bibnamefont {Song}}, \bibinfo {author} {\bibfnamefont {S.}~\bibnamefont {Yao}},\ and\ \bibinfo {author} {\bibfnamefont {Z.}~\bibnamefont {Wang}},\ }\bibfield  {title} {\bibinfo {title} {{Non-Hermitian Skin Effect} and {Chiral Damping} in {Open Quantum Systems}},\ }\href {https://doi.org/10.1103/PhysRevLett.123.170401} {\bibfield  {journal} {\bibinfo  {journal} {Phys. Rev. Lett.}\ }\textbf {\bibinfo {volume} {123}},\ \bibinfo {pages} {170401} (\bibinfo {year} {2019})}\BibitemShut {NoStop}%
\bibitem [{\citenamefont {Yoshida}\ \emph {et~al.}(2020)\citenamefont {Yoshida}, \citenamefont {Kudo}, \citenamefont {Katsura},\ and\ \citenamefont {Hatsugai}}]{PhysRevResearch.2.033428}%
  \BibitemOpen
  \bibfield  {author} {\bibinfo {author} {\bibfnamefont {T.}~\bibnamefont {Yoshida}}, \bibinfo {author} {\bibfnamefont {K.}~\bibnamefont {Kudo}}, \bibinfo {author} {\bibfnamefont {H.}~\bibnamefont {Katsura}},\ and\ \bibinfo {author} {\bibfnamefont {Y.}~\bibnamefont {Hatsugai}},\ }\bibfield  {title} {\bibinfo {title} {Fate of fractional quantum {Hall} states in open quantum systems: {Characterization} of correlated topological states for the full {Liouvillian}},\ }\href {https://doi.org/10.1103/PhysRevResearch.2.033428} {\bibfield  {journal} {\bibinfo  {journal} {Phys. Rev. Res.}\ }\textbf {\bibinfo {volume} {2}},\ \bibinfo {pages} {033428} (\bibinfo {year} {2020})}\BibitemShut {NoStop}%
\bibitem [{\citenamefont {Rudner}\ and\ \citenamefont {Levitov}(2009)}]{PhysRevLett.102.065703}%
  \BibitemOpen
  \bibfield  {author} {\bibinfo {author} {\bibfnamefont {M.~S.}\ \bibnamefont {Rudner}}\ and\ \bibinfo {author} {\bibfnamefont {L.~S.}\ \bibnamefont {Levitov}},\ }\bibfield  {title} {\bibinfo {title} {Topological {Transition} in a {Non}-{Hermitian} {Quantum} {Walk}},\ }\href {https://doi.org/10.1103/PhysRevLett.102.065703} {\bibfield  {journal} {\bibinfo  {journal} {Phys. Rev. Lett.}\ }\textbf {\bibinfo {volume} {102}},\ \bibinfo {pages} {065703} (\bibinfo {year} {2009})}\BibitemShut {NoStop}%
\bibitem [{\citenamefont {Malzard}\ \emph {et~al.}(2015)\citenamefont {Malzard}, \citenamefont {Poli},\ and\ \citenamefont {Schomerus}}]{PhysRevLett.115.200402}%
  \BibitemOpen
  \bibfield  {author} {\bibinfo {author} {\bibfnamefont {S.}~\bibnamefont {Malzard}}, \bibinfo {author} {\bibfnamefont {C.}~\bibnamefont {Poli}},\ and\ \bibinfo {author} {\bibfnamefont {H.}~\bibnamefont {Schomerus}},\ }\bibfield  {title} {\bibinfo {title} {Topologically {Protected Defect States} in {Open Photonic Systems} with {Non-Hermitian Charge-Conjugation} and {Parity-Time Symmetry}},\ }\href {https://doi.org/10.1103/PhysRevLett.115.200402} {\bibfield  {journal} {\bibinfo  {journal} {Phys. Rev. Lett.}\ }\textbf {\bibinfo {volume} {115}},\ \bibinfo {pages} {200402} (\bibinfo {year} {2015})}\BibitemShut {NoStop}%
\bibitem [{\citenamefont {Leykam}\ \emph {et~al.}(2017)\citenamefont {Leykam}, \citenamefont {Bliokh}, \citenamefont {Huang}, \citenamefont {Chong},\ and\ \citenamefont {Nori}}]{PhysRevLett.118.040401}%
  \BibitemOpen
  \bibfield  {author} {\bibinfo {author} {\bibfnamefont {D.}~\bibnamefont {Leykam}}, \bibinfo {author} {\bibfnamefont {K.~Y.}\ \bibnamefont {Bliokh}}, \bibinfo {author} {\bibfnamefont {C.}~\bibnamefont {Huang}}, \bibinfo {author} {\bibfnamefont {Y.~D.}\ \bibnamefont {Chong}},\ and\ \bibinfo {author} {\bibfnamefont {F.}~\bibnamefont {Nori}},\ }\bibfield  {title} {\bibinfo {title} {Edge {Modes}, {Degeneracies}, and {Topological Numbers} in {Non-Hermitian Systems}},\ }\href {https://doi.org/10.1103/PhysRevLett.118.040401} {\bibfield  {journal} {\bibinfo  {journal} {Phys. Rev. Lett.}\ }\textbf {\bibinfo {volume} {118}},\ \bibinfo {pages} {040401} (\bibinfo {year} {2017})}\BibitemShut {NoStop}%
\bibitem [{\citenamefont {Campos~Venuti}\ \emph {et~al.}(2017)\citenamefont {Campos~Venuti}, \citenamefont {Ma}, \citenamefont {Saleur},\ and\ \citenamefont {Haas}}]{PhysRevA.96.053858}%
  \BibitemOpen
  \bibfield  {author} {\bibinfo {author} {\bibfnamefont {L.}~\bibnamefont {Campos~Venuti}}, \bibinfo {author} {\bibfnamefont {Z.}~\bibnamefont {Ma}}, \bibinfo {author} {\bibfnamefont {H.}~\bibnamefont {Saleur}},\ and\ \bibinfo {author} {\bibfnamefont {S.}~\bibnamefont {Haas}},\ }\bibfield  {title} {\bibinfo {title} {Topological protection of coherence in a dissipative environment},\ }\href {https://doi.org/10.1103/PhysRevA.96.053858} {\bibfield  {journal} {\bibinfo  {journal} {Phys. Rev. A}\ }\textbf {\bibinfo {volume} {96}},\ \bibinfo {pages} {053858} (\bibinfo {year} {2017})}\BibitemShut {NoStop}%
\bibitem [{\citenamefont {Gong}\ \emph {et~al.}(2017)\citenamefont {Gong}, \citenamefont {Higashikawa},\ and\ \citenamefont {Ueda}}]{PhysRevLett.118.200401}%
  \BibitemOpen
  \bibfield  {author} {\bibinfo {author} {\bibfnamefont {Z.}~\bibnamefont {Gong}}, \bibinfo {author} {\bibfnamefont {S.}~\bibnamefont {Higashikawa}},\ and\ \bibinfo {author} {\bibfnamefont {M.}~\bibnamefont {Ueda}},\ }\bibfield  {title} {\bibinfo {title} {{Zeno Hall Effect}},\ }\href {https://doi.org/10.1103/PhysRevLett.118.200401} {\bibfield  {journal} {\bibinfo  {journal} {Phys. Rev. Lett.}\ }\textbf {\bibinfo {volume} {118}},\ \bibinfo {pages} {200401} (\bibinfo {year} {2017})}\BibitemShut {NoStop}%
\bibitem [{\citenamefont {Yao}\ and\ \citenamefont {Wang}(2018)}]{PhysRevLett.121.086803}%
  \BibitemOpen
  \bibfield  {author} {\bibinfo {author} {\bibfnamefont {S.}~\bibnamefont {Yao}}\ and\ \bibinfo {author} {\bibfnamefont {Z.}~\bibnamefont {Wang}},\ }\bibfield  {title} {\bibinfo {title} {Edge {States} and {Topological Invariants} of {Non-Hermitian Systems}},\ }\href {https://doi.org/10.1103/PhysRevLett.121.086803} {\bibfield  {journal} {\bibinfo  {journal} {Phys. Rev. Lett.}\ }\textbf {\bibinfo {volume} {121}},\ \bibinfo {pages} {086803} (\bibinfo {year} {2018})}\BibitemShut {NoStop}%
\bibitem [{\citenamefont {Kunst}\ \emph {et~al.}(2018)\citenamefont {Kunst}, \citenamefont {Edvardsson}, \citenamefont {Budich},\ and\ \citenamefont {Bergholtz}}]{PhysRevLett.121.026808}%
  \BibitemOpen
  \bibfield  {author} {\bibinfo {author} {\bibfnamefont {F.~K.}\ \bibnamefont {Kunst}}, \bibinfo {author} {\bibfnamefont {E.}~\bibnamefont {Edvardsson}}, \bibinfo {author} {\bibfnamefont {J.~C.}\ \bibnamefont {Budich}},\ and\ \bibinfo {author} {\bibfnamefont {E.~J.}\ \bibnamefont {Bergholtz}},\ }\bibfield  {title} {\bibinfo {title} {Biorthogonal {Bulk-Boundary Correspondence} in {Non-Hermitian Systems}},\ }\href {https://doi.org/10.1103/PhysRevLett.121.026808} {\bibfield  {journal} {\bibinfo  {journal} {Phys. Rev. Lett.}\ }\textbf {\bibinfo {volume} {121}},\ \bibinfo {pages} {026808} (\bibinfo {year} {2018})}\BibitemShut {NoStop}%
\bibitem [{\citenamefont {Gong}\ \emph {et~al.}(2018)\citenamefont {Gong}, \citenamefont {Ashida}, \citenamefont {Kawabata}, \citenamefont {Takasan}, \citenamefont {Higashikawa},\ and\ \citenamefont {Ueda}}]{PhysRevX.8.031079}%
  \BibitemOpen
  \bibfield  {author} {\bibinfo {author} {\bibfnamefont {Z.}~\bibnamefont {Gong}}, \bibinfo {author} {\bibfnamefont {Y.}~\bibnamefont {Ashida}}, \bibinfo {author} {\bibfnamefont {K.}~\bibnamefont {Kawabata}}, \bibinfo {author} {\bibfnamefont {K.}~\bibnamefont {Takasan}}, \bibinfo {author} {\bibfnamefont {S.}~\bibnamefont {Higashikawa}},\ and\ \bibinfo {author} {\bibfnamefont {M.}~\bibnamefont {Ueda}},\ }\bibfield  {title} {\bibinfo {title} {Topological {Phases} of {Non-Hermitian Systems}},\ }\href {https://doi.org/10.1103/PhysRevX.8.031079} {\bibfield  {journal} {\bibinfo  {journal} {Phys. Rev. X}\ }\textbf {\bibinfo {volume} {8}},\ \bibinfo {pages} {031079} (\bibinfo {year} {2018})}\BibitemShut {NoStop}%
\bibitem [{\citenamefont {Wang}\ and\ \citenamefont {Clerk}(2019)}]{PhysRevA.99.063834}%
  \BibitemOpen
  \bibfield  {author} {\bibinfo {author} {\bibfnamefont {Y.-X.}\ \bibnamefont {Wang}}\ and\ \bibinfo {author} {\bibfnamefont {A.~A.}\ \bibnamefont {Clerk}},\ }\bibfield  {title} {\bibinfo {title} {{Non-Hermitian} dynamics without dissipation in quantum systems},\ }\href {https://doi.org/10.1103/PhysRevA.99.063834} {\bibfield  {journal} {\bibinfo  {journal} {Phys. Rev. A}\ }\textbf {\bibinfo {volume} {99}},\ \bibinfo {pages} {063834} (\bibinfo {year} {2019})}\BibitemShut {NoStop}%
\bibitem [{\citenamefont {Kawabata}\ \emph {et~al.}(2019)\citenamefont {Kawabata}, \citenamefont {Shiozaki}, \citenamefont {Ueda},\ and\ \citenamefont {Sato}}]{PhysRevX.9.041015}%
  \BibitemOpen
  \bibfield  {author} {\bibinfo {author} {\bibfnamefont {K.}~\bibnamefont {Kawabata}}, \bibinfo {author} {\bibfnamefont {K.}~\bibnamefont {Shiozaki}}, \bibinfo {author} {\bibfnamefont {M.}~\bibnamefont {Ueda}},\ and\ \bibinfo {author} {\bibfnamefont {M.}~\bibnamefont {Sato}},\ }\bibfield  {title} {\bibinfo {title} {Symmetry and {Topology} in {Non-Hermitian Physics}},\ }\href {https://doi.org/10.1103/PhysRevX.9.041015} {\bibfield  {journal} {\bibinfo  {journal} {Phys. Rev. X}\ }\textbf {\bibinfo {volume} {9}},\ \bibinfo {pages} {041015} (\bibinfo {year} {2019})}\BibitemShut {NoStop}%
\bibitem [{\citenamefont {Borgnia}\ \emph {et~al.}(2020)\citenamefont {Borgnia}, \citenamefont {Kruchkov},\ and\ \citenamefont {Slager}}]{PhysRevLett.124.056802}%
  \BibitemOpen
  \bibfield  {author} {\bibinfo {author} {\bibfnamefont {D.~S.}\ \bibnamefont {Borgnia}}, \bibinfo {author} {\bibfnamefont {A.~J.}\ \bibnamefont {Kruchkov}},\ and\ \bibinfo {author} {\bibfnamefont {R.-J.}\ \bibnamefont {Slager}},\ }\bibfield  {title} {\bibinfo {title} {{Non-Hermitian Boundary Modes} and {Topology}},\ }\href {https://doi.org/10.1103/PhysRevLett.124.056802} {\bibfield  {journal} {\bibinfo  {journal} {Phys. Rev. Lett.}\ }\textbf {\bibinfo {volume} {124}},\ \bibinfo {pages} {056802} (\bibinfo {year} {2020})}\BibitemShut {NoStop}%
\bibitem [{\citenamefont {Ashida}\ \emph {et~al.}(2020)\citenamefont {Ashida}, \citenamefont {Gong},\ and\ \citenamefont {Ueda}}]{ashida2020non}%
  \BibitemOpen
  \bibfield  {author} {\bibinfo {author} {\bibfnamefont {Y.}~\bibnamefont {Ashida}}, \bibinfo {author} {\bibfnamefont {Z.}~\bibnamefont {Gong}},\ and\ \bibinfo {author} {\bibfnamefont {M.}~\bibnamefont {Ueda}},\ }\bibfield  {title} {\bibinfo {title} {Non-hermitian physics},\ }\href@noop {} {\bibfield  {journal} {\bibinfo  {journal} {Advances in Physics}\ }\textbf {\bibinfo {volume} {69}},\ \bibinfo {pages} {249} (\bibinfo {year} {2020})}\BibitemShut {NoStop}%
\bibitem [{\citenamefont {Carmichael}(2009)}]{tra1}%
  \BibitemOpen
  \bibfield  {author} {\bibinfo {author} {\bibfnamefont {H.}~\bibnamefont {Carmichael}},\ }\href@noop {} {\emph {\bibinfo {title} {An open systems approach to quantum optics: lectures presented at the Universit{\'e} Libre de Bruxelles, October 28 to November 4, 1991}}},\ Vol.~\bibinfo {volume} {18}\ (\bibinfo  {publisher} {Springer Science \& Business Media},\ \bibinfo {year} {2009})\BibitemShut {NoStop}%
\bibitem [{\citenamefont {Gneiting}\ \emph {et~al.}(2021)\citenamefont {Gneiting}, \citenamefont {Rozhkov},\ and\ \citenamefont {Nori}}]{tra2}%
  \BibitemOpen
  \bibfield  {author} {\bibinfo {author} {\bibfnamefont {C.}~\bibnamefont {Gneiting}}, \bibinfo {author} {\bibfnamefont {A.~V.}\ \bibnamefont {Rozhkov}},\ and\ \bibinfo {author} {\bibfnamefont {F.}~\bibnamefont {Nori}},\ }\bibfield  {title} {\bibinfo {title} {Jump-time unraveling of {Markovian} open quantum systems},\ }\href {https://doi.org/10.1103/PhysRevA.104.062212} {\bibfield  {journal} {\bibinfo  {journal} {Phys. Rev. A}\ }\textbf {\bibinfo {volume} {104}},\ \bibinfo {pages} {062212} (\bibinfo {year} {2021})}\BibitemShut {NoStop}%
\bibitem [{\citenamefont {Gneiting}\ \emph {et~al.}(2022)\citenamefont {Gneiting}, \citenamefont {Koottandavida}, \citenamefont {Rozhkov},\ and\ \citenamefont {Nori}}]{top2}%
  \BibitemOpen
  \bibfield  {author} {\bibinfo {author} {\bibfnamefont {C.}~\bibnamefont {Gneiting}}, \bibinfo {author} {\bibfnamefont {A.}~\bibnamefont {Koottandavida}}, \bibinfo {author} {\bibfnamefont {A.~V.}\ \bibnamefont {Rozhkov}},\ and\ \bibinfo {author} {\bibfnamefont {F.}~\bibnamefont {Nori}},\ }\bibfield  {title} {\bibinfo {title} {Unraveling the topology of dissipative quantum systems},\ }\href {https://doi.org/10.1103/PhysRevResearch.4.023036} {\bibfield  {journal} {\bibinfo  {journal} {Phys. Rev. Res.}\ }\textbf {\bibinfo {volume} {4}},\ \bibinfo {pages} {023036} (\bibinfo {year} {2022})}\BibitemShut {NoStop}%
\bibitem [{\citenamefont {Lu}\ \emph {et~al.}(2014)\citenamefont {Lu}, \citenamefont {Joannopoulos},\ and\ \citenamefont {Solja{\v{c}}i{\'{c}}}}]{Lu2014}%
  \BibitemOpen
  \bibfield  {author} {\bibinfo {author} {\bibfnamefont {L.}~\bibnamefont {Lu}}, \bibinfo {author} {\bibfnamefont {J.~D.}\ \bibnamefont {Joannopoulos}},\ and\ \bibinfo {author} {\bibfnamefont {M.}~\bibnamefont {Solja{\v{c}}i{\'{c}}}},\ }\bibfield  {title} {\bibinfo {title} {Topological photonics},\ }\href {https://doi.org/10.1038/nphoton.2014.248} {\bibfield  {journal} {\bibinfo  {journal} {Nature Photonics}\ }\textbf {\bibinfo {volume} {8}},\ \bibinfo {pages} {821} (\bibinfo {year} {2014})}\BibitemShut {NoStop}%
\bibitem [{\citenamefont {Ozawa}\ \emph {et~al.}(2019)\citenamefont {Ozawa}, \citenamefont {Price}, \citenamefont {Amo}, \citenamefont {Goldman}, \citenamefont {Hafezi}, \citenamefont {Lu}, \citenamefont {Rechtsman}, \citenamefont {Schuster}, \citenamefont {Simon}, \citenamefont {Zilberberg},\ and\ \citenamefont {Carusotto}}]{RevModPhys.91.015006}%
  \BibitemOpen
  \bibfield  {author} {\bibinfo {author} {\bibfnamefont {T.}~\bibnamefont {Ozawa}}, \bibinfo {author} {\bibfnamefont {H.~M.}\ \bibnamefont {Price}}, \bibinfo {author} {\bibfnamefont {A.}~\bibnamefont {Amo}}, \bibinfo {author} {\bibfnamefont {N.}~\bibnamefont {Goldman}}, \bibinfo {author} {\bibfnamefont {M.}~\bibnamefont {Hafezi}}, \bibinfo {author} {\bibfnamefont {L.}~\bibnamefont {Lu}}, \bibinfo {author} {\bibfnamefont {M.~C.}\ \bibnamefont {Rechtsman}}, \bibinfo {author} {\bibfnamefont {D.}~\bibnamefont {Schuster}}, \bibinfo {author} {\bibfnamefont {J.}~\bibnamefont {Simon}}, \bibinfo {author} {\bibfnamefont {O.}~\bibnamefont {Zilberberg}},\ and\ \bibinfo {author} {\bibfnamefont {I.}~\bibnamefont {Carusotto}},\ }\bibfield  {title} {\bibinfo {title} {Topological photonics},\ }\href {https://doi.org/10.1103/RevModPhys.91.015006} {\bibfield  {journal} {\bibinfo  {journal} {Rev. Mod. Phys.}\ }\textbf {\bibinfo {volume} {91}},\ \bibinfo {pages} {015006} (\bibinfo {year} {2019})}\BibitemShut {NoStop}%
\bibitem [{\citenamefont {Ghatak}\ and\ \citenamefont {Das}(2019)}]{Ghatak_2019}%
  \BibitemOpen
  \bibfield  {author} {\bibinfo {author} {\bibfnamefont {A.}~\bibnamefont {Ghatak}}\ and\ \bibinfo {author} {\bibfnamefont {T.}~\bibnamefont {Das}},\ }\bibfield  {title} {\bibinfo {title} {New topological invariants in non-{Hermitian} systems},\ }\href {https://doi.org/10.1088/1361-648X/ab11b3} {\bibfield  {journal} {\bibinfo  {journal} {Journal of Physics: Condensed Matter}\ }\textbf {\bibinfo {volume} {31}},\ \bibinfo {pages} {263001} (\bibinfo {year} {2019})}\BibitemShut {NoStop}%
\bibitem [{\citenamefont {Rider}\ \emph {et~al.}(2019)\citenamefont {Rider}, \citenamefont {Palmer}, \citenamefont {Pocock}, \citenamefont {Xiao}, \citenamefont {Arroyo~Huidobro},\ and\ \citenamefont {Giannini}}]{10.1063/1.5086433}%
  \BibitemOpen
  \bibfield  {author} {\bibinfo {author} {\bibfnamefont {M.~S.}\ \bibnamefont {Rider}}, \bibinfo {author} {\bibfnamefont {S.~J.}\ \bibnamefont {Palmer}}, \bibinfo {author} {\bibfnamefont {S.~R.}\ \bibnamefont {Pocock}}, \bibinfo {author} {\bibfnamefont {X.}~\bibnamefont {Xiao}}, \bibinfo {author} {\bibfnamefont {P.}~\bibnamefont {Arroyo~Huidobro}},\ and\ \bibinfo {author} {\bibfnamefont {V.}~\bibnamefont {Giannini}},\ }\bibfield  {title} {\bibinfo {title} {{A perspective on topological nanophotonics: {Current} status and future challenges}},\ }\href {https://doi.org/10.1063/1.5086433} {\bibfield  {journal} {\bibinfo  {journal} {Journal of Applied Physics}\ }\textbf {\bibinfo {volume} {125}},\ \bibinfo {pages} {120901} (\bibinfo {year} {2019})}\BibitemShut {NoStop}%
\bibitem [{\citenamefont {Smirnova}\ \emph {et~al.}(2020)\citenamefont {Smirnova}, \citenamefont {Leykam}, \citenamefont {Chong},\ and\ \citenamefont {Kivshar}}]{10.1063/1.5142397}%
  \BibitemOpen
  \bibfield  {author} {\bibinfo {author} {\bibfnamefont {D.}~\bibnamefont {Smirnova}}, \bibinfo {author} {\bibfnamefont {D.}~\bibnamefont {Leykam}}, \bibinfo {author} {\bibfnamefont {Y.}~\bibnamefont {Chong}},\ and\ \bibinfo {author} {\bibfnamefont {Y.}~\bibnamefont {Kivshar}},\ }\bibfield  {title} {\bibinfo {title} {{Nonlinear topological photonics}},\ }\href {https://doi.org/10.1063/1.5142397} {\bibfield  {journal} {\bibinfo  {journal} {Applied Physics Reviews}\ }\textbf {\bibinfo {volume} {7}},\ \bibinfo {pages} {021306} (\bibinfo {year} {2020})}\BibitemShut {NoStop}%
\bibitem [{\citenamefont {Rudner}\ and\ \citenamefont {Lindner}(2020)}]{Rudner2020}%
  \BibitemOpen
  \bibfield  {author} {\bibinfo {author} {\bibfnamefont {M.~S.}\ \bibnamefont {Rudner}}\ and\ \bibinfo {author} {\bibfnamefont {N.~H.}\ \bibnamefont {Lindner}},\ }\bibfield  {title} {\bibinfo {title} {Band structure engineering and non-equilibrium dynamics in {Floquet} topological insulators},\ }\href {https://doi.org/10.1038/s42254-020-0170-z} {\bibfield  {journal} {\bibinfo  {journal} {Nature Reviews Physics}\ }\textbf {\bibinfo {volume} {2}},\ \bibinfo {pages} {229} (\bibinfo {year} {2020})}\BibitemShut {NoStop}%
\bibitem [{\citenamefont {Wang}\ \emph {et~al.}(2021)\citenamefont {Wang}, \citenamefont {Dutt}, \citenamefont {Yang}, \citenamefont {Wojcik}, \citenamefont {Vučković},\ and\ \citenamefont {Fan}}]{doi:10.1126/science.abf6568}%
  \BibitemOpen
  \bibfield  {author} {\bibinfo {author} {\bibfnamefont {K.}~\bibnamefont {Wang}}, \bibinfo {author} {\bibfnamefont {A.}~\bibnamefont {Dutt}}, \bibinfo {author} {\bibfnamefont {K.~Y.}\ \bibnamefont {Yang}}, \bibinfo {author} {\bibfnamefont {C.~C.}\ \bibnamefont {Wojcik}}, \bibinfo {author} {\bibfnamefont {J.}~\bibnamefont {Vučković}},\ and\ \bibinfo {author} {\bibfnamefont {S.}~\bibnamefont {Fan}},\ }\bibfield  {title} {\bibinfo {title} {Generating arbitrary topological windings of a non-{Hermitian} band},\ }\href {https://doi.org/10.1126/science.abf6568} {\bibfield  {journal} {\bibinfo  {journal} {Science}\ }\textbf {\bibinfo {volume} {371}},\ \bibinfo {pages} {1240} (\bibinfo {year} {2021})}\BibitemShut {NoStop}%
\bibitem [{\citenamefont {Yang}\ \emph {et~al.}(2015)\citenamefont {Yang}, \citenamefont {Gao}, \citenamefont {Shi}, \citenamefont {Lin}, \citenamefont {Gao}, \citenamefont {Chong},\ and\ \citenamefont {Zhang}}]{PhysRevLett.114.114301}%
  \BibitemOpen
  \bibfield  {author} {\bibinfo {author} {\bibfnamefont {Z.}~\bibnamefont {Yang}}, \bibinfo {author} {\bibfnamefont {F.}~\bibnamefont {Gao}}, \bibinfo {author} {\bibfnamefont {X.}~\bibnamefont {Shi}}, \bibinfo {author} {\bibfnamefont {X.}~\bibnamefont {Lin}}, \bibinfo {author} {\bibfnamefont {Z.}~\bibnamefont {Gao}}, \bibinfo {author} {\bibfnamefont {Y.}~\bibnamefont {Chong}},\ and\ \bibinfo {author} {\bibfnamefont {B.}~\bibnamefont {Zhang}},\ }\bibfield  {title} {\bibinfo {title} {Topological {Acoustics}},\ }\href {https://doi.org/10.1103/PhysRevLett.114.114301} {\bibfield  {journal} {\bibinfo  {journal} {Phys. Rev. Lett.}\ }\textbf {\bibinfo {volume} {114}},\ \bibinfo {pages} {114301} (\bibinfo {year} {2015})}\BibitemShut {NoStop}%
\bibitem [{\citenamefont {Xue}\ \emph {et~al.}(2022)\citenamefont {Xue}, \citenamefont {Yang},\ and\ \citenamefont {Zhang}}]{Xue2022}%
  \BibitemOpen
  \bibfield  {author} {\bibinfo {author} {\bibfnamefont {H.}~\bibnamefont {Xue}}, \bibinfo {author} {\bibfnamefont {Y.}~\bibnamefont {Yang}},\ and\ \bibinfo {author} {\bibfnamefont {B.}~\bibnamefont {Zhang}},\ }\bibfield  {title} {\bibinfo {title} {Topological acoustics},\ }\href {https://doi.org/10.1038/s41578-022-00465-6} {\bibfield  {journal} {\bibinfo  {journal} {Nature Reviews Materials}\ }\textbf {\bibinfo {volume} {7}},\ \bibinfo {pages} {974} (\bibinfo {year} {2022})}\BibitemShut {NoStop}%
\bibitem [{\citenamefont {Huber}(2016)}]{Huber2016}%
  \BibitemOpen
  \bibfield  {author} {\bibinfo {author} {\bibfnamefont {S.~D.}\ \bibnamefont {Huber}},\ }\bibfield  {title} {\bibinfo {title} {Topological mechanics},\ }\href {https://doi.org/10.1038/nphys3801} {\bibfield  {journal} {\bibinfo  {journal} {Nature Physics}\ }\textbf {\bibinfo {volume} {12}},\ \bibinfo {pages} {621} (\bibinfo {year} {2016})}\BibitemShut {NoStop}%
\bibitem [{\citenamefont {Ni}\ \emph {et~al.}(2019)\citenamefont {Ni}, \citenamefont {Weiner}, \citenamefont {Al{\`u}},\ and\ \citenamefont {Khanikaev}}]{Ni2019}%
  \BibitemOpen
  \bibfield  {author} {\bibinfo {author} {\bibfnamefont {X.}~\bibnamefont {Ni}}, \bibinfo {author} {\bibfnamefont {M.}~\bibnamefont {Weiner}}, \bibinfo {author} {\bibfnamefont {A.}~\bibnamefont {Al{\`u}}},\ and\ \bibinfo {author} {\bibfnamefont {A.~B.}\ \bibnamefont {Khanikaev}},\ }\bibfield  {title} {\bibinfo {title} {Observation of higher-order topological acoustic states protected by generalized chiral symmetry},\ }\href {https://doi.org/10.1038/s41563-018-0252-9} {\bibfield  {journal} {\bibinfo  {journal} {Nature Materials}\ }\textbf {\bibinfo {volume} {18}},\ \bibinfo {pages} {113} (\bibinfo {year} {2019})}\BibitemShut {NoStop}%
\bibitem [{\citenamefont {Ma}\ \emph {et~al.}(2019)\citenamefont {Ma}, \citenamefont {Xiao},\ and\ \citenamefont {Chan}}]{Ma2019}%
  \BibitemOpen
  \bibfield  {author} {\bibinfo {author} {\bibfnamefont {G.}~\bibnamefont {Ma}}, \bibinfo {author} {\bibfnamefont {M.}~\bibnamefont {Xiao}},\ and\ \bibinfo {author} {\bibfnamefont {C.~T.}\ \bibnamefont {Chan}},\ }\bibfield  {title} {\bibinfo {title} {Topological phases in acoustic and mechanical systems},\ }\href {https://doi.org/10.1038/s42254-019-0030-x} {\bibfield  {journal} {\bibinfo  {journal} {Nature Reviews Physics}\ }\textbf {\bibinfo {volume} {1}},\ \bibinfo {pages} {281} (\bibinfo {year} {2019})}\BibitemShut {NoStop}%
\bibitem [{\citenamefont {Wintersperger}\ \emph {et~al.}(2020)\citenamefont {Wintersperger}, \citenamefont {Braun}, \citenamefont {{\"U}nal}, \citenamefont {Eckardt}, \citenamefont {Liberto}, \citenamefont {Goldman}, \citenamefont {Bloch},\ and\ \citenamefont {Aidelsburger}}]{Wintersperger2020}%
  \BibitemOpen
  \bibfield  {author} {\bibinfo {author} {\bibfnamefont {K.}~\bibnamefont {Wintersperger}}, \bibinfo {author} {\bibfnamefont {C.}~\bibnamefont {Braun}}, \bibinfo {author} {\bibfnamefont {F.~N.}\ \bibnamefont {{\"U}nal}}, \bibinfo {author} {\bibfnamefont {A.}~\bibnamefont {Eckardt}}, \bibinfo {author} {\bibfnamefont {M.~D.}\ \bibnamefont {Liberto}}, \bibinfo {author} {\bibfnamefont {N.}~\bibnamefont {Goldman}}, \bibinfo {author} {\bibfnamefont {I.}~\bibnamefont {Bloch}},\ and\ \bibinfo {author} {\bibfnamefont {M.}~\bibnamefont {Aidelsburger}},\ }\bibfield  {title} {\bibinfo {title} {Realization of an anomalous {Floquet} topological system with ultracold atoms},\ }\href {https://doi.org/10.1038/s41567-020-0949-y} {\bibfield  {journal} {\bibinfo  {journal} {Nature Physics}\ }\textbf {\bibinfo {volume} {16}},\ \bibinfo {pages} {1058} (\bibinfo {year} {2020})}\BibitemShut {NoStop}%
\bibitem [{\citenamefont {Li}\ \emph {et~al.}(2020)\citenamefont {Li}, \citenamefont {Lee},\ and\ \citenamefont {Gong}}]{PhysRevLett.124.250402}%
  \BibitemOpen
  \bibfield  {author} {\bibinfo {author} {\bibfnamefont {L.}~\bibnamefont {Li}}, \bibinfo {author} {\bibfnamefont {C.~H.}\ \bibnamefont {Lee}},\ and\ \bibinfo {author} {\bibfnamefont {J.}~\bibnamefont {Gong}},\ }\bibfield  {title} {\bibinfo {title} {{Topological Switch} for {Non-Hermitian Skin Effect} in {Cold-Atom Systems} with {Loss}},\ }\href {https://doi.org/10.1103/PhysRevLett.124.250402} {\bibfield  {journal} {\bibinfo  {journal} {Phys. Rev. Lett.}\ }\textbf {\bibinfo {volume} {124}},\ \bibinfo {pages} {250402} (\bibinfo {year} {2020})}\BibitemShut {NoStop}%
\bibitem [{\citenamefont {Semeghini}\ \emph {et~al.}(2021)\citenamefont {Semeghini}, \citenamefont {Levine}, \citenamefont {Keesling}, \citenamefont {Ebadi}, \citenamefont {Wang}, \citenamefont {Bluvstein}, \citenamefont {Verresen}, \citenamefont {Pichler}, \citenamefont {Kalinowski}, \citenamefont {Samajdar}, \citenamefont {Omran}, \citenamefont {Sachdev}, \citenamefont {Vishwanath}, \citenamefont {Greiner}, \citenamefont {Vuletić},\ and\ \citenamefont {Lukin}}]{doi:10.1126/science.abi8794}%
  \BibitemOpen
  \bibfield  {author} {\bibinfo {author} {\bibfnamefont {G.}~\bibnamefont {Semeghini}}, \bibinfo {author} {\bibfnamefont {H.}~\bibnamefont {Levine}}, \bibinfo {author} {\bibfnamefont {A.}~\bibnamefont {Keesling}}, \bibinfo {author} {\bibfnamefont {S.}~\bibnamefont {Ebadi}}, \bibinfo {author} {\bibfnamefont {T.~T.}\ \bibnamefont {Wang}}, \bibinfo {author} {\bibfnamefont {D.}~\bibnamefont {Bluvstein}}, \bibinfo {author} {\bibfnamefont {R.}~\bibnamefont {Verresen}}, \bibinfo {author} {\bibfnamefont {H.}~\bibnamefont {Pichler}}, \bibinfo {author} {\bibfnamefont {M.}~\bibnamefont {Kalinowski}}, \bibinfo {author} {\bibfnamefont {R.}~\bibnamefont {Samajdar}}, \bibinfo {author} {\bibfnamefont {A.}~\bibnamefont {Omran}}, \bibinfo {author} {\bibfnamefont {S.}~\bibnamefont {Sachdev}}, \bibinfo {author} {\bibfnamefont {A.}~\bibnamefont {Vishwanath}}, \bibinfo {author} {\bibfnamefont {M.}~\bibnamefont {Greiner}}, \bibinfo {author} {\bibfnamefont {V.}~\bibnamefont {Vuletić}},\ and\ \bibinfo {author} {\bibfnamefont
  {M.~D.}\ \bibnamefont {Lukin}},\ }\bibfield  {title} {\bibinfo {title} {Probing topological spin liquids on a programmable quantum simulator},\ }\href {https://doi.org/10.1126/science.abi8794} {\bibfield  {journal} {\bibinfo  {journal} {Science}\ }\textbf {\bibinfo {volume} {374}},\ \bibinfo {pages} {1242} (\bibinfo {year} {2021})}\BibitemShut {NoStop}%
\bibitem [{\citenamefont {Serra-Garcia}\ \emph {et~al.}(2019)\citenamefont {Serra-Garcia}, \citenamefont {S\"usstrunk},\ and\ \citenamefont {Huber}}]{PhysRevB.99.020304}%
  \BibitemOpen
  \bibfield  {author} {\bibinfo {author} {\bibfnamefont {M.}~\bibnamefont {Serra-Garcia}}, \bibinfo {author} {\bibfnamefont {R.}~\bibnamefont {S\"usstrunk}},\ and\ \bibinfo {author} {\bibfnamefont {S.~D.}\ \bibnamefont {Huber}},\ }\bibfield  {title} {\bibinfo {title} {Observation of quadrupole transitions and edge mode topology in an {LC} circuit network},\ }\href {https://doi.org/10.1103/PhysRevB.99.020304} {\bibfield  {journal} {\bibinfo  {journal} {Phys. Rev. B}\ }\textbf {\bibinfo {volume} {99}},\ \bibinfo {pages} {020304} (\bibinfo {year} {2019})}\BibitemShut {NoStop}%
\bibitem [{\citenamefont {Gilbert}(2021)}]{Gilbert2021}%
  \BibitemOpen
  \bibfield  {author} {\bibinfo {author} {\bibfnamefont {M.~J.}\ \bibnamefont {Gilbert}},\ }\bibfield  {title} {\bibinfo {title} {Topological electronics},\ }\href {https://doi.org/10.1038/s42005-021-00569-5} {\bibfield  {journal} {\bibinfo  {journal} {Communications Physics}\ }\textbf {\bibinfo {volume} {4}},\ \bibinfo {pages} {70} (\bibinfo {year} {2021})}\BibitemShut {NoStop}%
\bibitem [{\citenamefont {Kim}\ \emph {et~al.}(2021)\citenamefont {Kim}, \citenamefont {Zhang}, \citenamefont {Ferreira}, \citenamefont {Banker}, \citenamefont {Iverson}, \citenamefont {Sipahigil}, \citenamefont {Bello}, \citenamefont {Gonz\'alez-Tudela}, \citenamefont {Mirhosseini},\ and\ \citenamefont {Painter}}]{PhysRevX.11.011015}%
  \BibitemOpen
  \bibfield  {author} {\bibinfo {author} {\bibfnamefont {E.}~\bibnamefont {Kim}}, \bibinfo {author} {\bibfnamefont {X.}~\bibnamefont {Zhang}}, \bibinfo {author} {\bibfnamefont {V.~S.}\ \bibnamefont {Ferreira}}, \bibinfo {author} {\bibfnamefont {J.}~\bibnamefont {Banker}}, \bibinfo {author} {\bibfnamefont {J.~K.}\ \bibnamefont {Iverson}}, \bibinfo {author} {\bibfnamefont {A.}~\bibnamefont {Sipahigil}}, \bibinfo {author} {\bibfnamefont {M.}~\bibnamefont {Bello}}, \bibinfo {author} {\bibfnamefont {A.}~\bibnamefont {Gonz\'alez-Tudela}}, \bibinfo {author} {\bibfnamefont {M.}~\bibnamefont {Mirhosseini}},\ and\ \bibinfo {author} {\bibfnamefont {O.}~\bibnamefont {Painter}},\ }\bibfield  {title} {\bibinfo {title} {{Quantum Electrodynamics} in a {Topological Waveguide}},\ }\href {https://doi.org/10.1103/PhysRevX.11.011015} {\bibfield  {journal} {\bibinfo  {journal} {Phys. Rev. X}\ }\textbf {\bibinfo {volume} {11}},\ \bibinfo {pages} {011015} (\bibinfo {year} {2021})}\BibitemShut {NoStop}%
\bibitem [{\citenamefont {Gisin}(1984)}]{PhysRevLett.52.1657}%
  \BibitemOpen
  \bibfield  {author} {\bibinfo {author} {\bibfnamefont {N.}~\bibnamefont {Gisin}},\ }\bibfield  {title} {\bibinfo {title} {{Quantum Measurements} and {Stochastic Processes}},\ }\href {https://doi.org/10.1103/PhysRevLett.52.1657} {\bibfield  {journal} {\bibinfo  {journal} {Phys. Rev. Lett.}\ }\textbf {\bibinfo {volume} {52}},\ \bibinfo {pages} {1657} (\bibinfo {year} {1984})}\BibitemShut {NoStop}%
\bibitem [{\citenamefont {Diósi}(1986)}]{DIOSI1986451}%
  \BibitemOpen
  \bibfield  {author} {\bibinfo {author} {\bibfnamefont {L.}~\bibnamefont {Diósi}},\ }\bibfield  {title} {\bibinfo {title} {Stochastic pure state representation for open quantum systems},\ }\href {https://doi.org/https://doi.org/10.1016/0375-9601(86)90692-4} {\bibfield  {journal} {\bibinfo  {journal} {Physics Letters A}\ }\textbf {\bibinfo {volume} {114}},\ \bibinfo {pages} {451} (\bibinfo {year} {1986})}\BibitemShut {NoStop}%
\bibitem [{\citenamefont {Belavkin}(1990)}]{Belavkin1990}%
  \BibitemOpen
  \bibfield  {author} {\bibinfo {author} {\bibfnamefont {V.~P.}\ \bibnamefont {Belavkin}},\ }\bibfield  {title} {\bibinfo {title} {A stochastic posterior {Schr{\"o}dinger} equation for counting nondemolition measurement},\ }\href {https://doi.org/10.1007/BF00398273} {\bibfield  {journal} {\bibinfo  {journal} {Letters in Mathematical Physics}\ }\textbf {\bibinfo {volume} {20}},\ \bibinfo {pages} {85} (\bibinfo {year} {1990})}\BibitemShut {NoStop}%
\bibitem [{\citenamefont {Gardiner}\ \emph {et~al.}(1992)\citenamefont {Gardiner}, \citenamefont {Parkins},\ and\ \citenamefont {Zoller}}]{PhysRevA.46.4363}%
  \BibitemOpen
  \bibfield  {author} {\bibinfo {author} {\bibfnamefont {C.~W.}\ \bibnamefont {Gardiner}}, \bibinfo {author} {\bibfnamefont {A.~S.}\ \bibnamefont {Parkins}},\ and\ \bibinfo {author} {\bibfnamefont {P.}~\bibnamefont {Zoller}},\ }\bibfield  {title} {\bibinfo {title} {Wave-function quantum stochastic differential equations and quantum-jump simulation methods},\ }\href {https://doi.org/10.1103/PhysRevA.46.4363} {\bibfield  {journal} {\bibinfo  {journal} {Phys. Rev. A}\ }\textbf {\bibinfo {volume} {46}},\ \bibinfo {pages} {4363} (\bibinfo {year} {1992})}\BibitemShut {NoStop}%
\bibitem [{\citenamefont {Dalibard}\ \emph {et~al.}(1992)\citenamefont {Dalibard}, \citenamefont {Castin},\ and\ \citenamefont {M\o{}lmer}}]{PhysRevLett.68.580}%
  \BibitemOpen
  \bibfield  {author} {\bibinfo {author} {\bibfnamefont {J.}~\bibnamefont {Dalibard}}, \bibinfo {author} {\bibfnamefont {Y.}~\bibnamefont {Castin}},\ and\ \bibinfo {author} {\bibfnamefont {K.}~\bibnamefont {M\o{}lmer}},\ }\bibfield  {title} {\bibinfo {title} {Wave-function approach to dissipative processes in quantum optics},\ }\href {https://doi.org/10.1103/PhysRevLett.68.580} {\bibfield  {journal} {\bibinfo  {journal} {Phys. Rev. Lett.}\ }\textbf {\bibinfo {volume} {68}},\ \bibinfo {pages} {580} (\bibinfo {year} {1992})}\BibitemShut {NoStop}%
\bibitem [{\citenamefont {Gisin}\ and\ \citenamefont {Percival}(1992)}]{NGisin_1992}%
  \BibitemOpen
  \bibfield  {author} {\bibinfo {author} {\bibfnamefont {N.}~\bibnamefont {Gisin}}\ and\ \bibinfo {author} {\bibfnamefont {I.~C.}\ \bibnamefont {Percival}},\ }\bibfield  {title} {\bibinfo {title} {The quantum-state diffusion model applied to open systems},\ }\href {https://doi.org/10.1088/0305-4470/25/21/023} {\bibfield  {journal} {\bibinfo  {journal} {Journal of Physics A: Mathematical and General}\ }\textbf {\bibinfo {volume} {25}},\ \bibinfo {pages} {5677} (\bibinfo {year} {1992})}\BibitemShut {NoStop}%
\bibitem [{\citenamefont {Plenio}\ and\ \citenamefont {Knight}(1998)}]{RevModPhys.70.101}%
  \BibitemOpen
  \bibfield  {author} {\bibinfo {author} {\bibfnamefont {M.~B.}\ \bibnamefont {Plenio}}\ and\ \bibinfo {author} {\bibfnamefont {P.~L.}\ \bibnamefont {Knight}},\ }\bibfield  {title} {\bibinfo {title} {The quantum-jump approach to dissipative dynamics in quantum optics},\ }\href {https://doi.org/10.1103/RevModPhys.70.101} {\bibfield  {journal} {\bibinfo  {journal} {Rev. Mod. Phys.}\ }\textbf {\bibinfo {volume} {70}},\ \bibinfo {pages} {101} (\bibinfo {year} {1998})}\BibitemShut {NoStop}%
\bibitem [{\citenamefont {Su}\ \emph {et~al.}(1979)\citenamefont {Su}, \citenamefont {Schrieffer},\ and\ \citenamefont {Heeger}}]{top1}%
  \BibitemOpen
  \bibfield  {author} {\bibinfo {author} {\bibfnamefont {W.~P.}\ \bibnamefont {Su}}, \bibinfo {author} {\bibfnamefont {J.~R.}\ \bibnamefont {Schrieffer}},\ and\ \bibinfo {author} {\bibfnamefont {A.~J.}\ \bibnamefont {Heeger}},\ }\bibfield  {title} {\bibinfo {title} {Solitons in {Polyacetylene}},\ }\href {https://doi.org/10.1103/PhysRevLett.42.1698} {\bibfield  {journal} {\bibinfo  {journal} {Phys. Rev. Lett.}\ }\textbf {\bibinfo {volume} {42}},\ \bibinfo {pages} {1698} (\bibinfo {year} {1979})}\BibitemShut {NoStop}%
\bibitem [{\citenamefont {Nagourney}\ \emph {et~al.}(1986)\citenamefont {Nagourney}, \citenamefont {Sandberg},\ and\ \citenamefont {Dehmelt}}]{PhysRevLett.56.2797}%
  \BibitemOpen
  \bibfield  {author} {\bibinfo {author} {\bibfnamefont {W.}~\bibnamefont {Nagourney}}, \bibinfo {author} {\bibfnamefont {J.}~\bibnamefont {Sandberg}},\ and\ \bibinfo {author} {\bibfnamefont {H.}~\bibnamefont {Dehmelt}},\ }\bibfield  {title} {\bibinfo {title} {Shelved optical electron amplifier: {Observation} of quantum jumps},\ }\href {https://doi.org/10.1103/PhysRevLett.56.2797} {\bibfield  {journal} {\bibinfo  {journal} {Phys. Rev. Lett.}\ }\textbf {\bibinfo {volume} {56}},\ \bibinfo {pages} {2797} (\bibinfo {year} {1986})}\BibitemShut {NoStop}%
\bibitem [{\citenamefont {Sauter}\ \emph {et~al.}(1986)\citenamefont {Sauter}, \citenamefont {Neuhauser}, \citenamefont {Blatt},\ and\ \citenamefont {Toschek}}]{PhysRevLett.57.1696}%
  \BibitemOpen
  \bibfield  {author} {\bibinfo {author} {\bibfnamefont {T.}~\bibnamefont {Sauter}}, \bibinfo {author} {\bibfnamefont {W.}~\bibnamefont {Neuhauser}}, \bibinfo {author} {\bibfnamefont {R.}~\bibnamefont {Blatt}},\ and\ \bibinfo {author} {\bibfnamefont {P.~E.}\ \bibnamefont {Toschek}},\ }\bibfield  {title} {\bibinfo {title} {Observation of {Quantum Jumps}},\ }\href {https://doi.org/10.1103/PhysRevLett.57.1696} {\bibfield  {journal} {\bibinfo  {journal} {Phys. Rev. Lett.}\ }\textbf {\bibinfo {volume} {57}},\ \bibinfo {pages} {1696} (\bibinfo {year} {1986})}\BibitemShut {NoStop}%
\bibitem [{\citenamefont {Bergquist}\ \emph {et~al.}(1986)\citenamefont {Bergquist}, \citenamefont {Hulet}, \citenamefont {Itano},\ and\ \citenamefont {Wineland}}]{PhysRevLett.57.1699}%
  \BibitemOpen
  \bibfield  {author} {\bibinfo {author} {\bibfnamefont {J.~C.}\ \bibnamefont {Bergquist}}, \bibinfo {author} {\bibfnamefont {R.~G.}\ \bibnamefont {Hulet}}, \bibinfo {author} {\bibfnamefont {W.~M.}\ \bibnamefont {Itano}},\ and\ \bibinfo {author} {\bibfnamefont {D.~J.}\ \bibnamefont {Wineland}},\ }\bibfield  {title} {\bibinfo {title} {Observation of {Quantum Jumps} in a {Single Atom}},\ }\href {https://doi.org/10.1103/PhysRevLett.57.1699} {\bibfield  {journal} {\bibinfo  {journal} {Phys. Rev. Lett.}\ }\textbf {\bibinfo {volume} {57}},\ \bibinfo {pages} {1699} (\bibinfo {year} {1986})}\BibitemShut {NoStop}%
\bibitem [{\citenamefont {Peil}\ and\ \citenamefont {Gabrielse}(1999)}]{PhysRevLett.83.1287}%
  \BibitemOpen
  \bibfield  {author} {\bibinfo {author} {\bibfnamefont {S.}~\bibnamefont {Peil}}\ and\ \bibinfo {author} {\bibfnamefont {G.}~\bibnamefont {Gabrielse}},\ }\bibfield  {title} {\bibinfo {title} {Observing the {Quantum Limit} of an {Electron Cyclotron}: {QND Measurements} of {Quantum Jumps} between {Fock States}},\ }\href {https://doi.org/10.1103/PhysRevLett.83.1287} {\bibfield  {journal} {\bibinfo  {journal} {Phys. Rev. Lett.}\ }\textbf {\bibinfo {volume} {83}},\ \bibinfo {pages} {1287} (\bibinfo {year} {1999})}\BibitemShut {NoStop}%
\bibitem [{\citenamefont {Gustavsson}\ \emph {et~al.}(2006)\citenamefont {Gustavsson}, \citenamefont {Leturcq}, \citenamefont {Simovi\ifmmode~\check{c}\else \v{c}\fi{}}, \citenamefont {Schleser}, \citenamefont {Ihn}, \citenamefont {Studerus}, \citenamefont {Ensslin}, \citenamefont {Driscoll},\ and\ \citenamefont {Gossard}}]{PhysRevLett.96.076605}%
  \BibitemOpen
  \bibfield  {author} {\bibinfo {author} {\bibfnamefont {S.}~\bibnamefont {Gustavsson}}, \bibinfo {author} {\bibfnamefont {R.}~\bibnamefont {Leturcq}}, \bibinfo {author} {\bibfnamefont {B.}~\bibnamefont {Simovi\ifmmode~\check{c}\else \v{c}\fi{}}}, \bibinfo {author} {\bibfnamefont {R.}~\bibnamefont {Schleser}}, \bibinfo {author} {\bibfnamefont {T.}~\bibnamefont {Ihn}}, \bibinfo {author} {\bibfnamefont {P.}~\bibnamefont {Studerus}}, \bibinfo {author} {\bibfnamefont {K.}~\bibnamefont {Ensslin}}, \bibinfo {author} {\bibfnamefont {D.~C.}\ \bibnamefont {Driscoll}},\ and\ \bibinfo {author} {\bibfnamefont {A.~C.}\ \bibnamefont {Gossard}},\ }\bibfield  {title} {\bibinfo {title} {Counting {Statistics} of {Single Electron Transport} in a {Quantum Dot}},\ }\href {https://doi.org/10.1103/PhysRevLett.96.076605} {\bibfield  {journal} {\bibinfo  {journal} {Phys. Rev. Lett.}\ }\textbf {\bibinfo {volume} {96}},\ \bibinfo {pages} {076605} (\bibinfo {year} {2006})}\BibitemShut {NoStop}%
\bibitem [{\citenamefont {Fujisawa}\ \emph {et~al.}(2006)\citenamefont {Fujisawa}, \citenamefont {Hayashi}, \citenamefont {Tomita},\ and\ \citenamefont {Hirayama}}]{doi:10.1126/science.1126788}%
  \BibitemOpen
  \bibfield  {author} {\bibinfo {author} {\bibfnamefont {T.}~\bibnamefont {Fujisawa}}, \bibinfo {author} {\bibfnamefont {T.}~\bibnamefont {Hayashi}}, \bibinfo {author} {\bibfnamefont {R.}~\bibnamefont {Tomita}},\ and\ \bibinfo {author} {\bibfnamefont {Y.}~\bibnamefont {Hirayama}},\ }\bibfield  {title} {\bibinfo {title} {Bidirectional {Counting} of {Single Electrons}},\ }\href {https://doi.org/10.1126/science.1126788} {\bibfield  {journal} {\bibinfo  {journal} {Science}\ }\textbf {\bibinfo {volume} {312}},\ \bibinfo {pages} {1634} (\bibinfo {year} {2006})}\BibitemShut {NoStop}%
\bibitem [{\citenamefont {Gleyzes}\ \emph {et~al.}(2007)\citenamefont {Gleyzes}, \citenamefont {Kuhr}, \citenamefont {Guerlin}, \citenamefont {Bernu}, \citenamefont {Del{\'e}glise}, \citenamefont {Busk~Hoff}, \citenamefont {Brune}, \citenamefont {Raimond},\ and\ \citenamefont {Haroche}}]{Gleyzes2007}%
  \BibitemOpen
  \bibfield  {author} {\bibinfo {author} {\bibfnamefont {S.}~\bibnamefont {Gleyzes}}, \bibinfo {author} {\bibfnamefont {S.}~\bibnamefont {Kuhr}}, \bibinfo {author} {\bibfnamefont {C.}~\bibnamefont {Guerlin}}, \bibinfo {author} {\bibfnamefont {J.}~\bibnamefont {Bernu}}, \bibinfo {author} {\bibfnamefont {S.}~\bibnamefont {Del{\'e}glise}}, \bibinfo {author} {\bibfnamefont {U.}~\bibnamefont {Busk~Hoff}}, \bibinfo {author} {\bibfnamefont {M.}~\bibnamefont {Brune}}, \bibinfo {author} {\bibfnamefont {J.-M.}\ \bibnamefont {Raimond}},\ and\ \bibinfo {author} {\bibfnamefont {S.}~\bibnamefont {Haroche}},\ }\bibfield  {title} {\bibinfo {title} {Quantum jumps of light recording the birth and death of a photon in a cavity},\ }\href {https://doi.org/10.1038/nature05589} {\bibfield  {journal} {\bibinfo  {journal} {Nature}\ }\textbf {\bibinfo {volume} {446}},\ \bibinfo {pages} {297} (\bibinfo {year} {2007})}\BibitemShut {NoStop}%
\bibitem [{\citenamefont {Kubanek}\ \emph {et~al.}(2009)\citenamefont {Kubanek}, \citenamefont {Koch}, \citenamefont {Sames}, \citenamefont {Ourjoumtsev}, \citenamefont {Pinkse}, \citenamefont {Murr},\ and\ \citenamefont {Rempe}}]{Kubanek2009}%
  \BibitemOpen
  \bibfield  {author} {\bibinfo {author} {\bibfnamefont {A.}~\bibnamefont {Kubanek}}, \bibinfo {author} {\bibfnamefont {M.}~\bibnamefont {Koch}}, \bibinfo {author} {\bibfnamefont {C.}~\bibnamefont {Sames}}, \bibinfo {author} {\bibfnamefont {A.}~\bibnamefont {Ourjoumtsev}}, \bibinfo {author} {\bibfnamefont {P.~W.~H.}\ \bibnamefont {Pinkse}}, \bibinfo {author} {\bibfnamefont {K.}~\bibnamefont {Murr}},\ and\ \bibinfo {author} {\bibfnamefont {G.}~\bibnamefont {Rempe}},\ }\bibfield  {title} {\bibinfo {title} {Photon-by-photon feedback control of a single-atom trajectory},\ }\href {https://doi.org/10.1038/nature08563} {\bibfield  {journal} {\bibinfo  {journal} {Nature}\ }\textbf {\bibinfo {volume} {462}},\ \bibinfo {pages} {898} (\bibinfo {year} {2009})}\BibitemShut {NoStop}%
\bibitem [{\citenamefont {Neumann}\ \emph {et~al.}(2010)\citenamefont {Neumann}, \citenamefont {Beck}, \citenamefont {Steiner}, \citenamefont {Rempp}, \citenamefont {Fedder}, \citenamefont {Hemmer}, \citenamefont {Wrachtrup},\ and\ \citenamefont {Jelezko}}]{doi:10.1126/science.1189075}%
  \BibitemOpen
  \bibfield  {author} {\bibinfo {author} {\bibfnamefont {P.}~\bibnamefont {Neumann}}, \bibinfo {author} {\bibfnamefont {J.}~\bibnamefont {Beck}}, \bibinfo {author} {\bibfnamefont {M.}~\bibnamefont {Steiner}}, \bibinfo {author} {\bibfnamefont {F.}~\bibnamefont {Rempp}}, \bibinfo {author} {\bibfnamefont {H.}~\bibnamefont {Fedder}}, \bibinfo {author} {\bibfnamefont {P.~R.}\ \bibnamefont {Hemmer}}, \bibinfo {author} {\bibfnamefont {J.}~\bibnamefont {Wrachtrup}},\ and\ \bibinfo {author} {\bibfnamefont {F.}~\bibnamefont {Jelezko}},\ }\bibfield  {title} {\bibinfo {title} {{Single-Shot Readout} of a {Single Nuclear Spin}},\ }\href {https://doi.org/10.1126/science.1189075} {\bibfield  {journal} {\bibinfo  {journal} {Science}\ }\textbf {\bibinfo {volume} {329}},\ \bibinfo {pages} {542} (\bibinfo {year} {2010})}\BibitemShut {NoStop}%
\bibitem [{\citenamefont {Vijay}\ \emph {et~al.}(2011)\citenamefont {Vijay}, \citenamefont {Slichter},\ and\ \citenamefont {Siddiqi}}]{PhysRevLett.106.110502}%
  \BibitemOpen
  \bibfield  {author} {\bibinfo {author} {\bibfnamefont {R.}~\bibnamefont {Vijay}}, \bibinfo {author} {\bibfnamefont {D.~H.}\ \bibnamefont {Slichter}},\ and\ \bibinfo {author} {\bibfnamefont {I.}~\bibnamefont {Siddiqi}},\ }\bibfield  {title} {\bibinfo {title} {Observation of {Quantum Jumps} in a {Superconducting Artificial Atom}},\ }\href {https://doi.org/10.1103/PhysRevLett.106.110502} {\bibfield  {journal} {\bibinfo  {journal} {Phys. Rev. Lett.}\ }\textbf {\bibinfo {volume} {106}},\ \bibinfo {pages} {110502} (\bibinfo {year} {2011})}\BibitemShut {NoStop}%
\bibitem [{\citenamefont {Sayrin}\ \emph {et~al.}(2011)\citenamefont {Sayrin}, \citenamefont {Dotsenko}, \citenamefont {Zhou}, \citenamefont {Peaudecerf}, \citenamefont {Rybarczyk}, \citenamefont {Gleyzes}, \citenamefont {Rouchon}, \citenamefont {Mirrahimi}, \citenamefont {Amini}, \citenamefont {Brune}, \citenamefont {Raimond},\ and\ \citenamefont {Haroche}}]{Sayrin2011}%
  \BibitemOpen
  \bibfield  {author} {\bibinfo {author} {\bibfnamefont {C.}~\bibnamefont {Sayrin}}, \bibinfo {author} {\bibfnamefont {I.}~\bibnamefont {Dotsenko}}, \bibinfo {author} {\bibfnamefont {X.}~\bibnamefont {Zhou}}, \bibinfo {author} {\bibfnamefont {B.}~\bibnamefont {Peaudecerf}}, \bibinfo {author} {\bibfnamefont {T.}~\bibnamefont {Rybarczyk}}, \bibinfo {author} {\bibfnamefont {S.}~\bibnamefont {Gleyzes}}, \bibinfo {author} {\bibfnamefont {P.}~\bibnamefont {Rouchon}}, \bibinfo {author} {\bibfnamefont {M.}~\bibnamefont {Mirrahimi}}, \bibinfo {author} {\bibfnamefont {H.}~\bibnamefont {Amini}}, \bibinfo {author} {\bibfnamefont {M.}~\bibnamefont {Brune}}, \bibinfo {author} {\bibfnamefont {J.-M.}\ \bibnamefont {Raimond}},\ and\ \bibinfo {author} {\bibfnamefont {S.}~\bibnamefont {Haroche}},\ }\bibfield  {title} {\bibinfo {title} {Real-time quantum feedback prepares and stabilizes photon number states},\ }\href {https://doi.org/10.1038/nature10376} {\bibfield  {journal} {\bibinfo  {journal} {Nature}\ }\textbf {\bibinfo
  {volume} {477}},\ \bibinfo {pages} {73} (\bibinfo {year} {2011})}\BibitemShut {NoStop}%
\bibitem [{\citenamefont {Pla}\ \emph {et~al.}(2013)\citenamefont {Pla}, \citenamefont {Tan}, \citenamefont {Dehollain}, \citenamefont {Lim}, \citenamefont {Morton}, \citenamefont {Zwanenburg}, \citenamefont {Jamieson}, \citenamefont {Dzurak},\ and\ \citenamefont {Morello}}]{Pla2013}%
  \BibitemOpen
  \bibfield  {author} {\bibinfo {author} {\bibfnamefont {J.~J.}\ \bibnamefont {Pla}}, \bibinfo {author} {\bibfnamefont {K.~Y.}\ \bibnamefont {Tan}}, \bibinfo {author} {\bibfnamefont {J.~P.}\ \bibnamefont {Dehollain}}, \bibinfo {author} {\bibfnamefont {W.~H.}\ \bibnamefont {Lim}}, \bibinfo {author} {\bibfnamefont {J.~J.~L.}\ \bibnamefont {Morton}}, \bibinfo {author} {\bibfnamefont {F.~A.}\ \bibnamefont {Zwanenburg}}, \bibinfo {author} {\bibfnamefont {D.~N.}\ \bibnamefont {Jamieson}}, \bibinfo {author} {\bibfnamefont {A.~S.}\ \bibnamefont {Dzurak}},\ and\ \bibinfo {author} {\bibfnamefont {A.}~\bibnamefont {Morello}},\ }\bibfield  {title} {\bibinfo {title} {High-fidelity readout and control of a nuclear spin qubit in silicon},\ }\href {https://doi.org/10.1038/nature12011} {\bibfield  {journal} {\bibinfo  {journal} {Nature}\ }\textbf {\bibinfo {volume} {496}},\ \bibinfo {pages} {334} (\bibinfo {year} {2013})}\BibitemShut {NoStop}%
\bibitem [{\citenamefont {Minev}\ \emph {et~al.}(2019)\citenamefont {Minev}, \citenamefont {Mundhada}, \citenamefont {Shankar}, \citenamefont {Reinhold}, \citenamefont {Guti{\'e}rrez-J{\'a}uregui}, \citenamefont {Schoelkopf}, \citenamefont {Mirrahimi}, \citenamefont {Carmichael},\ and\ \citenamefont {Devoret}}]{Minev2019}%
  \BibitemOpen
  \bibfield  {author} {\bibinfo {author} {\bibfnamefont {Z.~K.}\ \bibnamefont {Minev}}, \bibinfo {author} {\bibfnamefont {S.~O.}\ \bibnamefont {Mundhada}}, \bibinfo {author} {\bibfnamefont {S.}~\bibnamefont {Shankar}}, \bibinfo {author} {\bibfnamefont {P.}~\bibnamefont {Reinhold}}, \bibinfo {author} {\bibfnamefont {R.}~\bibnamefont {Guti{\'e}rrez-J{\'a}uregui}}, \bibinfo {author} {\bibfnamefont {R.~J.}\ \bibnamefont {Schoelkopf}}, \bibinfo {author} {\bibfnamefont {M.}~\bibnamefont {Mirrahimi}}, \bibinfo {author} {\bibfnamefont {H.~J.}\ \bibnamefont {Carmichael}},\ and\ \bibinfo {author} {\bibfnamefont {M.~H.}\ \bibnamefont {Devoret}},\ }\bibfield  {title} {\bibinfo {title} {To catch and reverse a quantum jump mid-flight},\ }\href {https://doi.org/10.1038/s41586-019-1287-z} {\bibfield  {journal} {\bibinfo  {journal} {Nature}\ }\textbf {\bibinfo {volume} {570}},\ \bibinfo {pages} {200} (\bibinfo {year} {2019})}\BibitemShut {NoStop}%
\bibitem [{\citenamefont {Kurzmann}\ \emph {et~al.}(2019)\citenamefont {Kurzmann}, \citenamefont {Stegmann}, \citenamefont {Kerski}, \citenamefont {Schott}, \citenamefont {Ludwig}, \citenamefont {Wieck}, \citenamefont {K\"onig}, \citenamefont {Lorke},\ and\ \citenamefont {Geller}}]{PhysRevLett.122.247403}%
  \BibitemOpen
  \bibfield  {author} {\bibinfo {author} {\bibfnamefont {A.}~\bibnamefont {Kurzmann}}, \bibinfo {author} {\bibfnamefont {P.}~\bibnamefont {Stegmann}}, \bibinfo {author} {\bibfnamefont {J.}~\bibnamefont {Kerski}}, \bibinfo {author} {\bibfnamefont {R.}~\bibnamefont {Schott}}, \bibinfo {author} {\bibfnamefont {A.}~\bibnamefont {Ludwig}}, \bibinfo {author} {\bibfnamefont {A.~D.}\ \bibnamefont {Wieck}}, \bibinfo {author} {\bibfnamefont {J.}~\bibnamefont {K\"onig}}, \bibinfo {author} {\bibfnamefont {A.}~\bibnamefont {Lorke}},\ and\ \bibinfo {author} {\bibfnamefont {M.}~\bibnamefont {Geller}},\ }\bibfield  {title} {\bibinfo {title} {{Optical Detection} of {Single-Electron Tunneling} into a {Semiconductor Quantum Dot}},\ }\href {https://doi.org/10.1103/PhysRevLett.122.247403} {\bibfield  {journal} {\bibinfo  {journal} {Phys. Rev. Lett.}\ }\textbf {\bibinfo {volume} {122}},\ \bibinfo {pages} {247403} (\bibinfo {year} {2019})}\BibitemShut {NoStop}%
\bibitem [{\citenamefont {Gorini}\ \emph {et~al.}(2008)\citenamefont {Gorini}, \citenamefont {Kossakowski},\ and\ \citenamefont {Sudarshan}}]{master1}%
  \BibitemOpen
  \bibfield  {author} {\bibinfo {author} {\bibfnamefont {V.}~\bibnamefont {Gorini}}, \bibinfo {author} {\bibfnamefont {A.}~\bibnamefont {Kossakowski}},\ and\ \bibinfo {author} {\bibfnamefont {E.~C.~G.}\ \bibnamefont {Sudarshan}},\ }\bibfield  {title} {\bibinfo {title} {{Completely positive dynamical semigroups of N-level systems}},\ }\href {https://doi.org/10.1063/1.522979} {\bibfield  {journal} {\bibinfo  {journal} {Journal of Mathematical Physics}\ }\textbf {\bibinfo {volume} {17}},\ \bibinfo {pages} {821} (\bibinfo {year} {2008})}\BibitemShut {NoStop}%
\bibitem [{\citenamefont {Lindblad}(1976)}]{master2}%
  \BibitemOpen
  \bibfield  {author} {\bibinfo {author} {\bibfnamefont {G.}~\bibnamefont {Lindblad}},\ }\bibfield  {title} {\bibinfo {title} {On the generators of quantum dynamical semigroups},\ }\href {https://doi.org/10.1007/BF01608499} {\bibfield  {journal} {\bibinfo  {journal} {Communications in Mathematical Physics}\ }\textbf {\bibinfo {volume} {48}},\ \bibinfo {pages} {119} (\bibinfo {year} {1976})}\BibitemShut {NoStop}%
\bibitem [{\citenamefont {Mong}\ and\ \citenamefont {Shivamoggi}(2011)}]{10.1103/PhysRevB.83.125109}%
  \BibitemOpen
  \bibfield  {author} {\bibinfo {author} {\bibfnamefont {R.~S.~K.}\ \bibnamefont {Mong}}\ and\ \bibinfo {author} {\bibfnamefont {V.}~\bibnamefont {Shivamoggi}},\ }\bibfield  {title} {\bibinfo {title} {Edge states and the bulk-boundary correspondence in {D}irac hamiltonians},\ }\href {https://doi.org/10.1103/PhysRevB.83.125109} {\bibfield  {journal} {\bibinfo  {journal} {Phys. Rev. B}\ }\textbf {\bibinfo {volume} {83}},\ \bibinfo {pages} {125109} (\bibinfo {year} {2011})}\BibitemShut {NoStop}%
\bibitem [{\citenamefont {Li}\ \emph {et~al.}(2016)\citenamefont {Li}, \citenamefont {Yang},\ and\ \citenamefont {Chen}}]{10.1140/epjb/e2016-70325-x}%
  \BibitemOpen
  \bibfield  {author} {\bibinfo {author} {\bibfnamefont {L.}~\bibnamefont {Li}}, \bibinfo {author} {\bibfnamefont {C.}~\bibnamefont {Yang}},\ and\ \bibinfo {author} {\bibfnamefont {S.}~\bibnamefont {Chen}},\ }\bibfield  {title} {\bibinfo {title} {Topological invariants for phase transition points of one-dimensional {Z}2 topological systems},\ }\href {https://doi.org/10.1140/epjb/e2016-70325-x} {\bibfield  {journal} {\bibinfo  {journal} {Eur. Phys. J. B}\ }\textbf {\bibinfo {volume} {89}},\ \bibinfo {pages} {195} (\bibinfo {year} {2016})}\BibitemShut {NoStop}%
\bibitem [{\citenamefont {Niu}\ \emph {et~al.}(2021)\citenamefont {Niu}, \citenamefont {Yan}, \citenamefont {Zhou}, \citenamefont {Tao}, \citenamefont {Li}, \citenamefont {Liu}, \citenamefont {Zhang}, \citenamefont {Jia}, \citenamefont {Liu}, \citenamefont {Yan}, \citenamefont {Chen},\ and\ \citenamefont {Yu}}]{10.1016/j.scib.2021.02.035}%
  \BibitemOpen
  \bibfield  {author} {\bibinfo {author} {\bibfnamefont {J.}~\bibnamefont {Niu}}, \bibinfo {author} {\bibfnamefont {T.}~\bibnamefont {Yan}}, \bibinfo {author} {\bibfnamefont {Y.}~\bibnamefont {Zhou}}, \bibinfo {author} {\bibfnamefont {Z.}~\bibnamefont {Tao}}, \bibinfo {author} {\bibfnamefont {X.}~\bibnamefont {Li}}, \bibinfo {author} {\bibfnamefont {W.}~\bibnamefont {Liu}}, \bibinfo {author} {\bibfnamefont {L.}~\bibnamefont {Zhang}}, \bibinfo {author} {\bibfnamefont {H.}~\bibnamefont {Jia}}, \bibinfo {author} {\bibfnamefont {S.}~\bibnamefont {Liu}}, \bibinfo {author} {\bibfnamefont {Z.}~\bibnamefont {Yan}}, \bibinfo {author} {\bibfnamefont {Y.}~\bibnamefont {Chen}},\ and\ \bibinfo {author} {\bibfnamefont {D.}~\bibnamefont {Yu}},\ }\bibfield  {title} {\bibinfo {title} {Simulation of higher-order topological phases and related topological phase transitions in a superconducting qubit},\ }\href {https://doi.org/https://doi.org/10.1016/j.scib.2021.02.035} {\bibfield  {journal} {\bibinfo  {journal} {Science
  Bulletin}\ }\textbf {\bibinfo {volume} {66}},\ \bibinfo {pages} {1168} (\bibinfo {year} {2021})}\BibitemShut {NoStop}%
\bibitem [{\citenamefont {Johansson}\ \emph {et~al.}(2012)\citenamefont {Johansson}, \citenamefont {Nation},\ and\ \citenamefont {Nori}}]{qut1}%
  \BibitemOpen
  \bibfield  {author} {\bibinfo {author} {\bibfnamefont {J.}~\bibnamefont {Johansson}}, \bibinfo {author} {\bibfnamefont {P.}~\bibnamefont {Nation}},\ and\ \bibinfo {author} {\bibfnamefont {F.}~\bibnamefont {Nori}},\ }\bibfield  {title} {\bibinfo {title} {Qutip: An open-source {Python} framework for the dynamics of open quantum systems},\ }\href {https://doi.org/https://doi.org/10.1016/j.cpc.2012.02.021} {\bibfield  {journal} {\bibinfo  {journal} {Computer Physics Communications}\ }\textbf {\bibinfo {volume} {183}},\ \bibinfo {pages} {1760} (\bibinfo {year} {2012})}\BibitemShut {NoStop}%
\bibitem [{dat()}]{data}%
  \BibitemOpen
  \href {https://github.com/Innovation562/Simulating-monitoring-induced-topological-phase-transitions-with-small-systems.git} {\emph {\bibinfo {title} {\rm The codes and data for this manuscript can be accessed through GitHub.}}}\BibitemShut {Stop}%
\bibitem [{\citenamefont {You}\ \emph {et~al.}(2007)\citenamefont {You}, \citenamefont {Hu}, \citenamefont {Ashhab},\ and\ \citenamefont {Nori}}]{10.1103/PhysRevB.75.140515}%
  \BibitemOpen
  \bibfield  {author} {\bibinfo {author} {\bibfnamefont {J.~Q.}\ \bibnamefont {You}}, \bibinfo {author} {\bibfnamefont {X.}~\bibnamefont {Hu}}, \bibinfo {author} {\bibfnamefont {S.}~\bibnamefont {Ashhab}},\ and\ \bibinfo {author} {\bibfnamefont {F.}~\bibnamefont {Nori}},\ }\bibfield  {title} {\bibinfo {title} {Low-decoherence flux qubit},\ }\href {https://doi.org/10.1103/PhysRevB.75.140515} {\bibfield  {journal} {\bibinfo  {journal} {Phys. Rev. B}\ }\textbf {\bibinfo {volume} {75}},\ \bibinfo {pages} {140515(R)} (\bibinfo {year} {2007})}\BibitemShut {NoStop}%
\bibitem [{\citenamefont {Ye}\ \emph {et~al.}(2021)\citenamefont {Ye}, \citenamefont {Peng}, \citenamefont {Naghiloo}, \citenamefont {Cunningham},\ and\ \citenamefont {O’Brien}}]{10.1103/PhysRevLett.127.050502}%
  \BibitemOpen
  \bibfield  {author} {\bibinfo {author} {\bibfnamefont {Y.}~\bibnamefont {Ye}}, \bibinfo {author} {\bibfnamefont {K.}~\bibnamefont {Peng}}, \bibinfo {author} {\bibfnamefont {M.}~\bibnamefont {Naghiloo}}, \bibinfo {author} {\bibfnamefont {G.}~\bibnamefont {Cunningham}},\ and\ \bibinfo {author} {\bibfnamefont {K.~P.}\ \bibnamefont {O’Brien}},\ }\bibfield  {title} {\bibinfo {title} {Engineering purely nonlinear coupling between superconducting qubits using a quarton},\ }\href {https://doi.org/10.1103/PhysRevLett.127.050502} {\bibfield  {journal} {\bibinfo  {journal} {Phys. Rev. Lett.}\ }\textbf {\bibinfo {volume} {127}},\ \bibinfo {pages} {050502} (\bibinfo {year} {2021})}\BibitemShut {NoStop}%
\bibitem [{\citenamefont {Abdi}\ \emph {et~al.}(2015)\citenamefont {Abdi}, \citenamefont {Pernpeintner}, \citenamefont {Gross}, \citenamefont {Huebl},\ and\ \citenamefont {Hartmann}}]{10.1103/PhysRevLett.114.173602}%
  \BibitemOpen
  \bibfield  {author} {\bibinfo {author} {\bibfnamefont {M.}~\bibnamefont {Abdi}}, \bibinfo {author} {\bibfnamefont {M.}~\bibnamefont {Pernpeintner}}, \bibinfo {author} {\bibfnamefont {R.}~\bibnamefont {Gross}}, \bibinfo {author} {\bibfnamefont {H.}~\bibnamefont {Huebl}},\ and\ \bibinfo {author} {\bibfnamefont {M.~J.}\ \bibnamefont {Hartmann}},\ }\bibfield  {title} {\bibinfo {title} {Quantum state engineering with circuit electromechanical three-body interactions},\ }\href {https://doi.org/10.1103/PhysRevLett.114.173602} {\bibfield  {journal} {\bibinfo  {journal} {Phys. Rev. Lett.}\ }\textbf {\bibinfo {volume} {114}},\ \bibinfo {pages} {173602} (\bibinfo {year} {2015})}\BibitemShut {NoStop}%
\bibitem [{\citenamefont {Kounalakis}\ \emph {et~al.}(2019)\citenamefont {Kounalakis}, \citenamefont {Blanter},\ and\ \citenamefont {Steele}}]{10.1038/s41534-019-0219-y}%
  \BibitemOpen
  \bibfield  {author} {\bibinfo {author} {\bibfnamefont {M.}~\bibnamefont {Kounalakis}}, \bibinfo {author} {\bibfnamefont {Y.~M.}\ \bibnamefont {Blanter}},\ and\ \bibinfo {author} {\bibfnamefont {G.~A.}\ \bibnamefont {Steele}},\ }\bibfield  {title} {\bibinfo {title} {Synthesizing multi-phonon quantum superposition states using flux-mediated three-body interactions with superconducting qubits},\ }\href {https://doi.org/10.1038/s41534-019-0219-y} {\bibfield  {journal} {\bibinfo  {journal} {npj Quantum Information}\ }\textbf {\bibinfo {volume} {5}},\ \bibinfo {pages} {100} (\bibinfo {year} {2019})}\BibitemShut {NoStop}%
\bibitem [{\citenamefont {Menke}\ \emph {et~al.}(2022)\citenamefont {Menke}, \citenamefont {Banner}, \citenamefont {Bergamaschi}, \citenamefont {Di~Paolo}, \citenamefont {Veps\"al\"ainen}, \citenamefont {Weber}, \citenamefont {Winik}, \citenamefont {Melville}, \citenamefont {Niedzielski}, \citenamefont {Rosenberg}, \citenamefont {Serniak}, \citenamefont {Schwartz}, \citenamefont {Yoder}, \citenamefont {Aspuru-Guzik}, \citenamefont {Gustavsson}, \citenamefont {Grover}, \citenamefont {Hirjibehedin}, \citenamefont {Kerman},\ and\ \citenamefont {Oliver}}]{10.1103/PhysRevLett.129.220501}%
  \BibitemOpen
  \bibfield  {author} {\bibinfo {author} {\bibfnamefont {T.}~\bibnamefont {Menke}}, \bibinfo {author} {\bibfnamefont {W.~P.}\ \bibnamefont {Banner}}, \bibinfo {author} {\bibfnamefont {T.~R.}\ \bibnamefont {Bergamaschi}}, \bibinfo {author} {\bibfnamefont {A.}~\bibnamefont {Di~Paolo}}, \bibinfo {author} {\bibfnamefont {A.}~\bibnamefont {Veps\"al\"ainen}}, \bibinfo {author} {\bibfnamefont {S.~J.}\ \bibnamefont {Weber}}, \bibinfo {author} {\bibfnamefont {R.}~\bibnamefont {Winik}}, \bibinfo {author} {\bibfnamefont {A.}~\bibnamefont {Melville}}, \bibinfo {author} {\bibfnamefont {B.~M.}\ \bibnamefont {Niedzielski}}, \bibinfo {author} {\bibfnamefont {D.}~\bibnamefont {Rosenberg}}, \bibinfo {author} {\bibfnamefont {K.}~\bibnamefont {Serniak}}, \bibinfo {author} {\bibfnamefont {M.~E.}\ \bibnamefont {Schwartz}}, \bibinfo {author} {\bibfnamefont {J.~L.}\ \bibnamefont {Yoder}}, \bibinfo {author} {\bibfnamefont {A.}~\bibnamefont {Aspuru-Guzik}}, \bibinfo {author} {\bibfnamefont {S.}~\bibnamefont {Gustavsson}}, \bibinfo
  {author} {\bibfnamefont {J.~A.}\ \bibnamefont {Grover}}, \bibinfo {author} {\bibfnamefont {C.~F.}\ \bibnamefont {Hirjibehedin}}, \bibinfo {author} {\bibfnamefont {A.~J.}\ \bibnamefont {Kerman}},\ and\ \bibinfo {author} {\bibfnamefont {W.~D.}\ \bibnamefont {Oliver}},\ }\bibfield  {title} {\bibinfo {title} {Demonstration of tunable three-body interactions between superconducting qubits},\ }\href {https://doi.org/10.1103/PhysRevLett.129.220501} {\bibfield  {journal} {\bibinfo  {journal} {Phys. Rev. Lett.}\ }\textbf {\bibinfo {volume} {129}},\ \bibinfo {pages} {220501} (\bibinfo {year} {2022})}\BibitemShut {NoStop}%
\bibitem [{\citenamefont {Teufel}\ \emph {et~al.}(2011)\citenamefont {Teufel}, \citenamefont {Donner}, \citenamefont {Li}, \citenamefont {Harlow}, \citenamefont {Allman}, \citenamefont {Cicak}, \citenamefont {Sirois}, \citenamefont {Whittaker}, \citenamefont {Lehnert},\ and\ \citenamefont {Simmonds}}]{10.1038/nature10261}%
  \BibitemOpen
  \bibfield  {author} {\bibinfo {author} {\bibfnamefont {J.~D.}\ \bibnamefont {Teufel}}, \bibinfo {author} {\bibfnamefont {T.}~\bibnamefont {Donner}}, \bibinfo {author} {\bibfnamefont {D.}~\bibnamefont {Li}}, \bibinfo {author} {\bibfnamefont {J.~W.}\ \bibnamefont {Harlow}}, \bibinfo {author} {\bibfnamefont {M.~S.}\ \bibnamefont {Allman}}, \bibinfo {author} {\bibfnamefont {K.}~\bibnamefont {Cicak}}, \bibinfo {author} {\bibfnamefont {A.~J.}\ \bibnamefont {Sirois}}, \bibinfo {author} {\bibfnamefont {J.~D.}\ \bibnamefont {Whittaker}}, \bibinfo {author} {\bibfnamefont {K.~W.}\ \bibnamefont {Lehnert}},\ and\ \bibinfo {author} {\bibfnamefont {R.~W.}\ \bibnamefont {Simmonds}},\ }\bibfield  {title} {\bibinfo {title} {Sideband cooling of micromechanical motion to the quantum ground state},\ }\href {https://doi.org/10.1038/nature10261} {\bibfield  {journal} {\bibinfo  {journal} {Nature}\ }\textbf {\bibinfo {volume} {475}},\ \bibinfo {pages} {359} (\bibinfo {year} {2011})}\BibitemShut {NoStop}%
\bibitem [{\citenamefont {Oberli}\ \emph {et~al.}(1990)\citenamefont {Oberli}, \citenamefont {Shah}, \citenamefont {Damen}, \citenamefont {Kuo}, \citenamefont {Henry}, \citenamefont {Lary},\ and\ \citenamefont {Goodnick}}]{10.1063/1.102525}%
  \BibitemOpen
  \bibfield  {author} {\bibinfo {author} {\bibfnamefont {D.~Y.}\ \bibnamefont {Oberli}}, \bibinfo {author} {\bibfnamefont {J.}~\bibnamefont {Shah}}, \bibinfo {author} {\bibfnamefont {T.~C.}\ \bibnamefont {Damen}}, \bibinfo {author} {\bibfnamefont {J.~M.}\ \bibnamefont {Kuo}}, \bibinfo {author} {\bibfnamefont {J.~E.}\ \bibnamefont {Henry}}, \bibinfo {author} {\bibfnamefont {J.}~\bibnamefont {Lary}},\ and\ \bibinfo {author} {\bibfnamefont {S.~M.}\ \bibnamefont {Goodnick}},\ }\bibfield  {title} {\bibinfo {title} {Optical phonon‐assisted tunneling in double quantum well structures},\ }\href {https://doi.org/10.1063/1.102525} {\bibfield  {journal} {\bibinfo  {journal} {Applied Physics Letters}\ }\textbf {\bibinfo {volume} {56}},\ \bibinfo {pages} {1239} (\bibinfo {year} {1990})}\BibitemShut {NoStop}%
\bibitem [{\citenamefont {Blaser}\ \emph {et~al.}(2001)\citenamefont {Blaser}, \citenamefont {Diehl}, \citenamefont {Beck}, \citenamefont {Faist}, \citenamefont {Oesterle}, \citenamefont {Xu}, \citenamefont {Barbieri},\ and\ \citenamefont {Beltram}}]{10.1109/3.910456}%
  \BibitemOpen
  \bibfield  {author} {\bibinfo {author} {\bibfnamefont {S.}~\bibnamefont {Blaser}}, \bibinfo {author} {\bibfnamefont {L.}~\bibnamefont {Diehl}}, \bibinfo {author} {\bibfnamefont {M.}~\bibnamefont {Beck}}, \bibinfo {author} {\bibfnamefont {J.}~\bibnamefont {Faist}}, \bibinfo {author} {\bibfnamefont {U.}~\bibnamefont {Oesterle}}, \bibinfo {author} {\bibfnamefont {J.}~\bibnamefont {Xu}}, \bibinfo {author} {\bibfnamefont {S.}~\bibnamefont {Barbieri}},\ and\ \bibinfo {author} {\bibfnamefont {F.}~\bibnamefont {Beltram}},\ }\bibfield  {title} {\bibinfo {title} {Characterization and modeling of quantum cascade lasers based on a photon-assisted tunneling transition},\ }\href {https://doi.org/10.1109/3.910456} {\bibfield  {journal} {\bibinfo  {journal} {IEEE Journal of Quantum Electronics}\ }\textbf {\bibinfo {volume} {37}},\ \bibinfo {pages} {448} (\bibinfo {year} {2001})}\BibitemShut {NoStop}%
\bibitem [{\citenamefont {Platero}\ and\ \citenamefont {Aguado}(2004)}]{10.1016/j.physrep.2004.01.004}%
  \BibitemOpen
  \bibfield  {author} {\bibinfo {author} {\bibfnamefont {G.}~\bibnamefont {Platero}}\ and\ \bibinfo {author} {\bibfnamefont {R.}~\bibnamefont {Aguado}},\ }\bibfield  {title} {\bibinfo {title} {Photon-assisted transport in semiconductor nanostructures},\ }\href {https://doi.org/https://doi.org/10.1016/j.physrep.2004.01.004} {\bibfield  {journal} {\bibinfo  {journal} {Physics Reports}\ }\textbf {\bibinfo {volume} {395}},\ \bibinfo {pages} {1} (\bibinfo {year} {2004})}\BibitemShut {NoStop}%
\bibitem [{\citenamefont {Hei}\ \emph {et~al.}(2023)\citenamefont {Hei}, \citenamefont {Li}, \citenamefont {Pan},\ and\ \citenamefont {Nori}}]{PhysRevLett.130.073602}%
  \BibitemOpen
  \bibfield  {author} {\bibinfo {author} {\bibfnamefont {X.-L.}\ \bibnamefont {Hei}}, \bibinfo {author} {\bibfnamefont {P.-B.}\ \bibnamefont {Li}}, \bibinfo {author} {\bibfnamefont {X.-F.}\ \bibnamefont {Pan}},\ and\ \bibinfo {author} {\bibfnamefont {F.}~\bibnamefont {Nori}},\ }\bibfield  {title} {\bibinfo {title} {Enhanced tripartite interactions in spin-magnon-mechanical hybrid systems},\ }\href {https://doi.org/10.1103/PhysRevLett.130.073602} {\bibfield  {journal} {\bibinfo  {journal} {Phys. Rev. Lett.}\ }\textbf {\bibinfo {volume} {130}},\ \bibinfo {pages} {073602} (\bibinfo {year} {2023})}\BibitemShut {NoStop}%
\bibitem [{\citenamefont {Wiseman}\ and\ \citenamefont {Milburn}(2009)}]{tra3}%
  \BibitemOpen
  \bibfield  {author} {\bibinfo {author} {\bibfnamefont {H.~M.}\ \bibnamefont {Wiseman}}\ and\ \bibinfo {author} {\bibfnamefont {G.~J.}\ \bibnamefont {Milburn}},\ }\href@noop {} {\emph {\bibinfo {title} {Quantum measurement and control}}}\ (\bibinfo  {publisher} {Cambridge university press},\ \bibinfo {year} {2009})\BibitemShut {NoStop}%
\end{thebibliography}%
\end{document}